\documentclass[aps,pre,preprint,groupedaddress,showpacs]{revtex4-1}
\usepackage{graphicx,color}
\usepackage{dcolumn}
\usepackage{bm}
\begin{document}

\title{Phase transition in the Bayesian estimation
\\
 of the default portfolio
}

\author{Masato Hisakado}
\email{hisakadom@yahoo.co.jp} 
\affiliation{
* Nomura Holdings, Inc., Otemachi 2-2-2, Chiyoda-ku, Tokyo 100-8130, Japan} 

\author{Shintaro Mori}
\email{shintaro.mori@hirosaki-u.ac.jp}
\affiliation{
\dag Department of Mathematics and Physics,
Graduate School of Science and Technology, 
Hirosaki University \\
Bunkyo-cho 3, Hirosaki, Aomori 036-8561, Japan}

\date{\today}

\begin{abstract}
  The probability of default (PD) estimation is an important process for financial institutions.
  The difficulty of the estimation depends on the correlations between borrowers. In this paper,
  we introduce a hierarchical Bayesian estimation method using the beta binomial distribution and consider a multi-year case with a temporal correlation. 
  A phase transition occurs when the temporal correlation decays
  by power decay. When the power index is less than one, the PD estimator
 does not converge. It is difficult to estimate the PD with limited historical data.
 Conversely, when the power index is greater than one, the convergence is the same as that of the binomial distribution.
 We provide a condition for the estimation of the PD and 
discuss the universality class of the phase transition.
We investigate the empirical default data history of rating agencies 
and their Fourier transformations to confirm the
the form of the correlation decay. The power spectrum of the decay history seems to be 1/f, which corresponds to a long memory.
But the estimated power index is much greater than one. If we collect adequate historical data,
the parameters can be estimated correctly.
\end{abstract}


\maketitle

\section{Introduction}

Anomalous diffusion is an emerging subject in many fields \cite{Bro,W2,G,M}.
The models describing such phenomena depend on long memory.
These are related to the phase transition, which
has received considerable interest in sociophysics \cite{galam,galam2} 
and econophysics \cite{Man}.
In previous papers, we investigated voting models that were similar
to the Keynesian beauty contest
\cite{Mori,Hisakado2,Hisakado3,Hisakado4,Hisakado5}.
This model has two kinds of phase transitions.
One is the information cascade transition, which is similar to the
 phase transition of the Ising model \cite{Hisakado3}.
The other is the convergence transition of
super-normal diffusion \cite{Hod,Hisakado2}.

Estimations of the probability of default (PD) and default correlation have been obtained from empirical studies on the historical data from credit events. These two parameters are important for pricing financial
products such as synthetic CDOs \cite{M2010,M2008,Sch}. Also called ``long run PDs", these parameters are important to financial institutions for portfolio management. If the number of defaults is minimal,
it is not easy to estimate these parameters \cite{Tas,FSA}.

In this paper, we introduce a Bayesian estimation method using the
beta-binomial distribution \cite{Hisakado,Wit}. For the usual cases the Merton
model, which incorporates the default correlation by 
the correlation of the asset price movements (asset correlation),
is used to estimate the PD and default correlation \cite{Mer}.
Monte Carlo simulations are necessary to estimate these
parameters, except in the limit of 
large homogeneous portfolios, where the Merton model is used
\cite{Sch}. In the beta-binomial case, default correlation,
instead of asset correlation, is used \cite{Hisakado}.
Moreover, we consider a multi-year case with temporal correlation,
which refers to a time-dependent correlation\cite{Tas,FSA}.

A phase transition occurs when the temporal correlation decays by
power-law. A power-law decay implies that the PD has a long memory 
compared to that of exponential decay \cite{Man}. When the
power index is less than one,
the estimator distribution of the PD
does not converge to the delta function.
Alternatively, when the power index is greater than one, the
convergence is the same as that of the normal case.
When the distribution does not converge, it is difficult to estimate the 
PD with limited data.
The required condition for estimating the PD is clarified. 
The critical exponents for the power-law decay of the correlation function
depend on microscopic features of the model.
The universality class of the phase transition is different
from those of the nonlinear P\'{o}lya urns \cite{Mori5,Mori6}.

To confirm the decay form of the temporal correlation,
we investigate the empirical default
data history using Fourier transformations. We determine whether the power
spectrum of the default history follows 1/f
\cite{Kes,Man}. When this condition is satisfied, it corresponds to the correlation
of the PD with long memory where a phase transition of the convergence exists.
However, it is difficult to accurately confirm a 1/f power spectrum when the estimation of the power index is much greater than one.
It follows that when there is adequate historical data,
parameters such as PD, default correlation, and temporal correlation can be estimated correctly.


The remainder of this paper is organized as follows.
In section 2, we introduce a hierarchical Bayesian estimation
method using the beta-binomial distribution. 
In section 3, we consider the convergence of the PD estimator. 
In section 4, we study the phase transition of the
 P\'{o}lya urn with a discount factor 
using an analytic method and a finite-size scaling analysis.
In section 5, we apply the Bayesian estimation to the empirical
data of default history.
Finally, the conclusions are presented in section 6.

\section{Bayesian Estimation using Beta-binomial distribution }

We denote the PD estimation as $\theta$ and default correlation as $ \rho_D$, where $0\leq\theta\leq1$ and $0\leq\rho_D\leq1$.
The distribution of $\theta$ and $\rho_D$ is $P(\theta,\rho_D)$. The number of obligors in the portfolio is $n$.
$\theta$ and $\rho_D$ are estimated using a Bayesian estimation.
We consider the Bernoulli random variables $X_i(i=1,2,\cdots,n)$ that take the values $1$ or $0$ .
When the obligor, $i$, is the default (non-default), $X_i=1(0)$.
 We define $X=\sum_{j=1}^{n} X_j$ and consider a default correlation for $X_i$, and not an asset correlation.

When the number of defaults is $k$, 
the Bayes formula for the posterior distribution $P(\theta,\rho_D|X=k)$ is
\begin{equation}
P(\theta,\rho_D|X=k)=
\frac{P(\theta,\rho_D,X=k)}{P(X=k)}
=\frac{P(X=k|\theta,\rho_D)f(\theta, \rho_D)}{P(X=k)},
\label{1}
\end{equation}
where $f(\theta, \rho_D)$ is the prior distribution.

We use the beta-binomial distribution for $P(X=k|\theta,\rho_D)$. 
The posterior distribution is given by
\begin{eqnarray}
P(\theta,\rho_D|X=k)&\propto& \frac{n!}{k!(n-k)!}
\frac{B(\alpha+k,n+\beta-k)}{B(\alpha,\beta)}f(\theta,\rho_D)
\nonumber \\
&\propto&\frac{\Gamma(\alpha+k)}{\Gamma(\alpha)}
\frac{\Gamma(n+\beta-k)}{\Gamma(\beta)}
\frac{\Gamma(\alpha+\beta)}{\Gamma(\alpha+\beta+n)}f(\theta,\rho_D),
\label{B}
\end{eqnarray}
where, 
$\theta=\frac{\alpha}{\alpha+\beta}$ and $\rho_D=\frac{1}{\alpha+\beta+1}$.
Hence, we obtain the relations $\alpha=\theta\frac{1-\rho_D}{\rho_D}$ and $\beta=(1-\theta)\frac{1-\rho_D}{\rho_D}$.
Here, we use the beta function $B(\alpha, \beta)=\Gamma(\alpha)\Gamma(\beta)/\Gamma(\alpha+\beta)$.

We consider the maximum \textit{a posteriori} (MAP) estimation of Eq. (\ref{B}). When the prior function $f(\theta,\rho_D)$ is a constant function, the maximum point is
\begin{eqnarray}
\frac{\partial P(\theta,\rho_D|X=k)}{\partial \theta}&\propto&
\frac{(1-\rho_D)}{\rho_D}\frac{\Gamma(\alpha+k)}{\Gamma(\alpha)}\frac{\Gamma(n+\beta-k)}{\Gamma(\beta)}(\varphi(\alpha+k)-\varphi(\alpha)-\varphi(\beta+n-k)+\varphi(\beta))
\nonumber \\
&=&
\frac{(1-\rho_D)}{\rho_D}\frac{\Gamma(\alpha+k)}{\Gamma(\alpha)}\frac{\Gamma(n+\beta-k)}{\Gamma(\beta)}(\sum_{i=1}^{k}\frac{1}{\alpha+i-1}-\sum_{i=1}^{n-k}\frac{1}{\beta+i-1})=0,
\end{eqnarray}
where $\varphi(x)$ is the digamma function.
The summation from $i=1$ to $k$ is a monotonously decreasing
function of $\theta$ because $\alpha$ increases, while the second summation in Eq. 3 is a monotonously increasing
function about $\theta$ because $\beta$ decreases.
When $\theta\sim 0$, the difference of the two summations is positive.
Conversely, when $\theta\sim 1$, the difference of the two summations becomes negative.
Hence, the function $P(\theta|X=k,\rho_D)$ has one peak in the range $0<\theta<1$. The multi-term case is provided in Appendix A.

Next, we consider the variable $\rho_D$. The maximum point is
\begin{eqnarray}
\frac{\partial P(\theta,z|X=k)}{\partial z}&\propto&\frac{\Gamma(\alpha+k)}{\Gamma(\alpha)}\frac{\Gamma(N+\beta-k)}{\Gamma(\beta)}
\frac{\Gamma(\alpha+\beta)}{\Gamma(\alpha+\beta+n)}
\nonumber \\
& &
(\sum_{i=1}^{k}\frac{\theta}{\theta z+i-1}+\sum_{i=1}^{k}\frac{1-\theta}{(1-\theta) z+i-1}
-\sum_{i=1}^{k}\frac{1}{z+i-1})=0,
\nonumber \\
\end{eqnarray}
 where $z=(1-\rho_D)/\rho_D$.

All the summations in the last term with parenthesis are monotonously decreasing functions about $z$. 
When $z\sim0$, the last term becomes positive.
 Conversely, when $z>>1$, the last term becomes 0.
When $(k-1)/n\leq\theta\leq k/n$ or is adequately close to this condition, the last term becomes positive.
$(k-1)/\theta\leq n-1$ and $(n-k-1)/(1-\theta)\leq n-1$ become $(k-1)/(n-1)\leq\theta\leq k/ (n-1)$.
In this case, the last term increases monotonously, and the peak is $z=\infty$ and $\rho_D=0$.
This implies that the optimization of $\rho_D$ is zero for the single term model. 
When $\theta$ is not adequately close to $(k-1)/n\leq\theta\leq k/n$, the last term changes from positive to negative as $z$ increases.
Therefore, one peak occurs in $P(\theta, z|X=k)$.

 We extend this method to the multi-year case.
 There are $n_i$ obligors in year $i$ and $k_i$ defaults occur. 
The prior distribution for the second year is the posterior distribution, which is calculated from the first year’s data.
In this way, the posterior distribution is updated every year.
We write the posterior distribution $P(\theta,\rho_{D}|k_1,k_2)$ as
\begin{equation}
P(\theta,\rho_D|k_1,k_2)
=\frac{P(k_2|\theta,\rho_D,k_1)}{P(k_2)}\frac{P(k_1|\theta,\rho_D)f(\theta, \rho_D)}{P(k_1)}.
\label{2}
\end{equation}

It is natural to assume that the number of defaults of the current year is affected by the
number of defaults in previous years, thus the defaults have a temporal correlation.
When the default rate is high (low), it is reasonable to
assume that the default rate will be high (low) in the next year.
This is similar to volatility clustering,
which has a long memory \cite{AR,GA} as well. 
We confirm this using empirical data in the following sections.

We introduce the temporal correlation by adjusting $\alpha$ and $\beta$, and  consider the $j$ th year. 
The number of obligors and defaults in the $j$ th year are $n_j$ and $k_j$.
In the same year, the correlation is $\rho_D$.
We set the temporal correlation parameters between the $i$ and $j$ th years; $d_{i-j}$ and $j<i$.
$\alpha $ and $\beta$ are adjusted to $\alpha+\sum_{j=1}^{i-1} d_{i-j}k_{j}$
and $\beta+\sum_{j=1}^{i-1} d_{i-j}(n_j-k_{j})$ \cite{Mori18}.
This implies that the previous years' data affects the present defaults.
It is easy to confirm that $d_i=1$ indicates that all the data is correlated to $\rho_D$.
When $d_i=0$, the data is independent each year.

\section{ Correlation Decay}

In the previous section, $d_i$ was introduced to represent the temporal correlation. In this section, to clarify the behavior of the parameter $d_i$, where $i=1,2,\cdots, T$ and $d_0=1$, the variance of the stochastic process is considered.
In each year the diffusion has $n_i$ steps and $k_i$ defaults, where  $i=1,2,\cdots,T$.

The adjustments related to parameters $\alpha$ and $\beta$ are the effects of the temporal correlation from the previous conclusions.
We shrink the previous years' conclusions and add them to the initial parameters for the adjustment process.
The shrinking ratio for the interval $i$ is $d_i$

The two term model is examined first.
We consider the relation between the first and second years.
$n_1$ and $k_1$ are the number of obligors and defaults, respectively, in the first year. 
The second year’s parameters become
$\alpha+d_1 k_1$ and $\beta+d_1 (n_1-k_1)$.
We consider the shrinking processes
from $\alpha$ to $\alpha+d_1 k_1$ and 
$\beta$ to $\beta+d_1 (n_1-k_1)$.
The variance of the second term of  process is 
$n_1d_1 pq+d_1 n_1(n_1-1)pq \rho_D$, where  $q=1-p$;
that is, we  approximate
$d_1 B_{\alpha,\beta}(k_1,n_1-k_1)\sim B_{\alpha,\beta}(d_1 k_1,d_1 n_1- d_1 k_1)  $
where $B_{\alpha,\beta}$ is the beta-binomial distribution with
parameters $\alpha$ and $\beta$.
We approximate this variance by  
$n_1 d_1 pq+d_1n_1(d_1n_1-1)pq \rho_D$, and
the difference becomes
$ n_1^2pq \rho_D d_1(1-d_1)\geq 0$.
Hence, the approximation is exact
when $d_1=0,1$ or $\rho_D=0$.
However, if $d_1\sim 0,1$ or $\rho_D\sim 0$,
this approximation can be used. 
In other cases, the real variance is larger than the approximation.
We use this approximation to study the meaning of this process.

For the defaults of the obligors, the hypothesis $d_1\sim 0$ or $1$ and $\rho_D\sim 0$ is given.
In other words, the temporal correlation is either a high or low case, or
a low correlation case.
Hereafter, we use this approximation to calculate the 
variance of this process.

We extend the stochastic process to the multi-year case.
Let $\{U_{t};t\ge 1\}$ be an independent and identically distributed (i.i.d.) sequence that is uniformly distributed on [0,1]. The discrete dynamics of the process is described by:
\begin{equation}
X(t+1)=\textbf{1}_{U_{t+1}\le Z_d (t)},
\end{equation}
when $n_i+1\leq t\leq n_{i+1}$.
Here $Z_d(t)$ is given by
\begin{equation}
Z_d (t)\equiv 
\frac{\alpha+\sum_{s=n_i}^{t}X(s)+\sum_{j=1}^{i}d_{i-j} k_j}{\alpha+\beta+(t-n_i)+\sum_{j=1}^{i}d_{i-j} n_j}.
\label{polyage}
\end{equation}

The expectation value of $X(t)$ is
$
\mbox{E}(X(t))=\alpha/(\alpha+\beta).
$
When $d_i=1$, the process is beta-binomial.

We consider the relationship between the year $i$ and $i+1$.
The distribution of year $i$ is a beta-binomial distribution. 
Hence, the conditional variance, $V_{i+1}$, of the year $i+1$ can be evaluated, using the above approximation, as
\begin{eqnarray}
V_{i+1}&\sim&\sum_{j=1}^{i+1}n_j d_{i+1-j}pq+(\sum_{j=1}^{i+1}n_j d_{i+1-j} )(\sum_{j=1}^{j+1}n_j d_{i+1-j} -1)pq\rho_D
\nonumber \\
&&-\sum_{j=1}^{i}n_jd_{i+1-j}pq -(\sum_{j=1}^{i}n_j  d_{i+1-j})(\sum_{j=1}^{i}n_j d_{i+1-j}-1)pq\rho_D
\nonumber \\
&=& pqn_{i+1}+pqn_{i+1}(n_{i+1}-1)\rho_D +2pq\rho_D n_{i+1}\sum_{j=1}^{i}n_j d_{i+1-j}.
\end{eqnarray}
Therefore, the difference of the summations $\sum_{j=1}^{i+1}n_jd_{i+1-j}$ and $\sum_{j=1}^{i}n_j d_{i+1-j}$ correspond to the 
 the variance of $(i+1)^{th}$ step. 
Therefore, using this approximation, the correlation between the $i^{th}$ and $j^{th}$ years is approximated by $\rho_D d_{i-j}$.
The term $d_{i-j}$ plays the role of a discount factor in the correlation $\rho_D$.
It can be seen that as time progresses, the correlation is discounted.
It is reasonable to assume a monotonically decreasing function for $d_i$ because the effects decrease as the distance between $i$ and $j$ increase. 
 
The total variance for the diffusion is approximated by
\begin{equation}
V\sim\sum_{i=1}^{T}pqn_{i}+\sum_{i=1}^{T}pqn_i(n_i-1)\rho_D+
2pq\rho_D\sum_{i>j}^{T}n_i n_j d_{i-j}.
\end{equation}
The first, second, and third terms correspond to the variance for binomial distribution,
constant correlation $\rho_D$ in the portfolio, and temporal correlation, respectively.

In summary, when $d_i\sim 0,1$ or $\rho_D\sim 0$, the correlation between year $i$ and year $j$ is approximated by
\[
  Corr \sim\rho_D \left(
    \begin{array}{ccccc}
     1 &  d_1 & d_2 &\cdots& d_T \\
   d_1      &  1 &  d_1 & \ddots&\vdots \\
  \ddots & \ddots & \ddots  & \ddots&\ddots  \\
 \vdots & \ddots & \ddots  & \ddots &d_1 \\
    d_T&  \cdots &  d_2&  d_1  & 1  \\
    \end{array}
  \right).
\]
The average PD, correlation of the Bernoulli random variables, and temporal correlation using this approximation are $p$, $\rho_D$, and $d_i$, respectively.

In the Bayesian estimation, if the scaled variance converges as the data increases, these parameters can be estimated correctly.
Conversely, if the variance does not converge, the parameters cannot be estimated.
It should also be considered whether the process has a stationary solution, which will be discussed regarding the spectrum analysis in the following sections.
 
It is difficult to estimate all the $d_i$ values due to limited data.
By introducing a prior distribution for $d_i$, the estimation becomes a hierarchical Bayesian estimation.
It is reasonable to assume that the prior distribution is a monotonically decreasing function.
Therefore, we considered two hyperprior distributions, an exponential and power decay, to have long memory.

\section{Phase transition in the estimation of PD}

In this section we determine whether the PD in the Bayesian estimation 
converges.
To simplify the model, we set $n_j= 1,j\ge 1$ in Eq. (\ref{polyage}).
This does not affect the outcome of the PD estimation.
Let $\{U_{t};t\ge 1\}$ be an independent and
identically distributed (i.i.d.) sequence that is uniformly distributed on [0,1]. 
The discrete dynamics of the process is described by:
\[
X(t+1)=\textbf{1}_{U_{t+1}\le Z_{d}(t)}.
\]
Here, $Z_{d}(t)$ is the weighted sum of $X(s),s\le t$ with the discount factor
$d_{t-s}$,
\begin{equation}
Z_{d}(t)\equiv 
\frac{\alpha+\sum_{s=1}^{t}X(s)d_{t-s}}{\alpha+\beta+\sum_{s=1}^{t}d_{t-s}}.
\label{polya}
\end{equation}
This is the P\'{o}lya urn model\cite{Polya} with a discount factor $\{d_i\}$.

The expectation value of $X(t)$ is
$
\mbox{E}(X(t))=\alpha/(\alpha+\beta).
$
The PD estimator is $Z(t)$,
\[
Z(t)\equiv \sum_{s=1}^{t}X(s)/t.
\]
The success of the PD estimation depends on the
the behavior of the variance of $Z(t)$. More specifically, if the variance of $Z(t)$ converges, then the PD can be estimated.

\subsection{Stochastic differential equation} 
First, the stochastic process is rewritten using 
$c_1 (t)=\sum_{s=1}^{t}X(s)$;
\begin{eqnarray}
c_1(t)&=&k \rightarrow k+1:
 P_{k,t}=\frac{\alpha+\sum_{s=1}^{t}X(s)d_{t-s}}{\alpha+\beta+\sum_{s=1}^{t}d_{t-s}},
\nonumber \\
c_1 (t)&=&k   \rightarrow k:
 Q_{k,t}=1-P_{k,t},
\label{pd}
\end{eqnarray}
where $P_{k,t}$ and $Q_{k,t}$ are the process probabilities.
The sum of $P_{k,t}$ and $Q_{k,t}$ is $1$.

For convenience, we define a new variable $\Delta_t$ such that
\begin{equation}
\Delta_t=2c_1(t)-t.
\label{d}
\end{equation}
We change the variables from $k$ to $\Delta_t$ and $X(s)$ to $x_s=2X(s)-1$. Given $\Delta_t=u$, we obtain a random walk model:
\begin{eqnarray}
\Delta_t&=&u \rightarrow u+1  :P_{u,t}=\frac{\alpha+\sum_{s=1}^{t}d_{t-s}(x_s+1)/2 }{\alpha+\beta+\sum_{s=1}^{t}d_{t-s}},
\nonumber \\
\Delta_t&=&u \rightarrow u-1  :Q_{u,t:s,t-r}=1-P_{u,t}.
\nonumber
\end{eqnarray}
We now consider the continuous limit $\epsilon \rightarrow 0$,
\begin{eqnarray}
Y_{\tau}&=&\epsilon\Delta_{[t/\epsilon]},
\nonumber \\
P(y,\tau)&=&\epsilon P(\Delta_t/\epsilon,t/\epsilon),
\end{eqnarray}
where $\tau=t/\epsilon$ and $y=\Delta_t/\epsilon$.
On approaching the continuous limit, we obtain the following stochastic partial differential equation:
\begin{equation}
\textrm{d}Y_{\tau}=
\frac{\alpha-\beta+\int_{\sigma=1}^{\tau}d(\tau-\sigma)\textrm{d}Y_{\sigma}}{\alpha+\beta+\int_{1}^{\tau}d(\tau-\sigma)\textrm{d}\sigma}
\textrm{d}\tau+\sqrt{\epsilon},
\label{ito13}
\end{equation}
where $d(\tau)$ is the continuous function of $d_t$, the discount factor, and $dY_\tau=\epsilon x_{[t/\epsilon]}$.

We are interested in the behavior of $Y_{\tau}$ in the limit $\tau\rightarrow \infty$.
We assume that the stationary solution is
\begin{equation}
Y_\infty=\bar{v}\tau,
\label{h3}
\end{equation}
where $\bar{v}$ is a constant.
Substituting Eq. (\ref{h3})into Eq. (\ref{ito13}), we obtain
\begin{equation}
\bar{v}=
\frac{\alpha-\beta+\bar{v}\hat{T}}{\alpha+\beta+\hat{T}},
\label{i3}
\end{equation}
where $\hat{T}=\lim_{\tau\rightarrow \infty} \int_{1}^{\tau}d(\tau-\sigma)\textrm{d}\sigma$.

Eq. (\ref{i3}) is a self-consistent equation.
When $\hat{T}<\infty$,  
Eq. (\ref{i3}) is solved when 
$\bar{v}=(\alpha-\beta)/(\alpha+\beta)$.
The process converges to the average point.
On the other hand,
when $\hat{T}\rightarrow\infty$, we can obtain the identity equation 
$\bar{v}=\bar{v}$, suggesting that the process does not converge
to the delta function.
The expected value of $Y_s$ is $(\alpha-\beta)/(\alpha+\beta)$.
Hence, the phase transition at the point $\hat{T}$
diverges to infinity.
When the distribution does not converge, we cannot estimate the parameters correctly, even if the amount of data increases.
This is a critical issue when using the Bayesian estimation.
In other words, $\hat{T}<\infty$ is a compulsory condition for parameter estimation.

\subsection{Correlation function and finite size scaling analysis}
To understand the phase transition, 
we investigated the correlation function, $C(t)$. $C(t)$ is defined as the correlation between $X(1)$ and $X(t)$ such that
\begin{equation}
C(t)\equiv \mbox{E}(X(t+1)|X(1)=1)-\mbox{E}(X(t+1)|X(1)=0)=
\frac{\mbox{Cov}(X(1),X(t+1))}{\mbox{V}(X(1))}.
\label{Cor}
\end{equation}
The function $C(t)$ represents the propagation of the memory of $X(1)$ to later variables $X(t+1)$. To understand the relationship between the variances of $Z(t)$ and $C(t)$, the variance of $Z(t)$ can be written as
\begin{equation}
\mbox{V}(Z(t))=\mbox{E}_{X(1)}(\mbox{V}(Z(t)|X(1)))
+\mbox{E}_{X(1)}((\mbox{E}(Z(t)|X(1))-\mbox{E}(Z(t)))^2, 
\label{eq:decomp}
\end{equation}
where $\mbox{V}(Z(t)|X(1))$ is the conditional variance of $Z(t)$ on $X(1)$.
The expectation value of $x$ is $\mbox{E}_{X(1)}(x)$ and the
probability function is $P(X(1))$. The second term on the right-hand side of  Eq. (\ref{eq:decomp}) represents the variance of $\mbox{E}(Z(t)|X(1))$ from the dependence on $X(1)$. In Eq. (\ref{eq:decomp}),
the second term is related to $C(t)$ as it
originates from the dependence of $\mbox{E}(Z(t)|X(1))$ on $X(1)$.
We write the second term of $C(t)$ as
\[
\mbox{E}_{X(1)}((\mbox{E}(Z(t)|X(1))-\mbox{E}(Z(t)))^2)=
\frac{1}{t^2}\frac{\alpha\beta}{(\alpha+\beta)^2}
 \left(\sum_{s=0}^{t-1}C(s)\right)^2.
\]
If $c=\lim_{t\to \infty}C(t)>0$, $\lim_{t\to\infty}\mbox{V}(Z(t))>0$
and $Z(t)$ does not converge.

Using $X(t+1)=\textbf{1}_{U_{t+1}\le Z_{d}(t)}$, we obtain the next relation
for the conditional expectation value of $X(t+1)$ with the condition $X(1)=x$,
$\mbox{E}(X(t+1)|X(1)=x)$, as
\[
\mbox{E}(X(t+1)|X(1)=x)
=\frac{\alpha+
\sum_{s=1}^{t}\mbox{E}(X(s)|X(1)=x)d_{t-s}}{\alpha+\beta+\sum_{s=1}^{t}d_{t-s}}.
\]
As $C(t)=\mbox{E}(X(t+1)|1)-\mbox{E}(X(t+1)|0)$, we obtain the following
recursive relation for $C(t)$ as
\begin{equation}
C(t)=\frac{\sum_{s=1}^{t}C(s-1)d_{t-s}}{\alpha+\beta+\sum_{s=1}^{t}d_{t-s}}. 
\label{eq:self}
\end{equation}
This recursive relation contains all information regarding the
asymptotic behavior of $C(t)$.
If one assumes a functional form
for $d_{i}$ with the initial condition $C(0)=1$, we can estimate 
$C(t)$ for $t\ge 1$.

\subsubsection{Exponential decay case}
We consider the exponential decay case, $d_{i}=r^{i},r\le 1$.
$\hat{T}$ is finite and there is no phase transition.
We decompose the numerator of Eq. (\ref{eq:self}) as
$C(t-1)+\sum_{s=1}^{t-1}C(s-1)r^{t-s}$. We rewrite the second
term using Eq. (\ref{eq:self}) for $t-1$ as
\[
\sum_{s=1}^{t-1}C(s-1)r^{t-s}=r\sum_{s=1}^{t-1}C(s-1)r^{t-1-s}
=r\cdot C(t-1)(\alpha+\beta+\sum_{s=1}^{t-1}r^{t-1-s}).
\]
We then obtain the next recursive relation for $C(t)$:
\begin{equation}
C(t)=
\frac{1+r(\alpha+\beta+\sum_{s=1}^{t-1}r^{t-1-s})}
     {\alpha+\beta+\sum_{s=1}^{t}r^{t-s}}C(t-1) .
\end{equation}

As we are interested in the asymptotic behavior of $C(t)$, we estimate
the decay rate, $r_{eff}$, with $C(t)\sim r_{eff}^{t}$, which gives 
\begin{equation}
r_{eff}\equiv \lim_{t\to \infty}C(t)/C(t-1)=r+\frac{1-r}{(\alpha+\beta)(1-r)+1}<1,
\end{equation}
where $r_{eff}<1$ for $r<1$, and $C(t)$ decays exponentially.

\begin{figure}[h]
\begin{center}
\begin{tabular}{c}
\begin{minipage}{0.5\hsize}
\begin{center}
\includegraphics[clip, width=8cm]{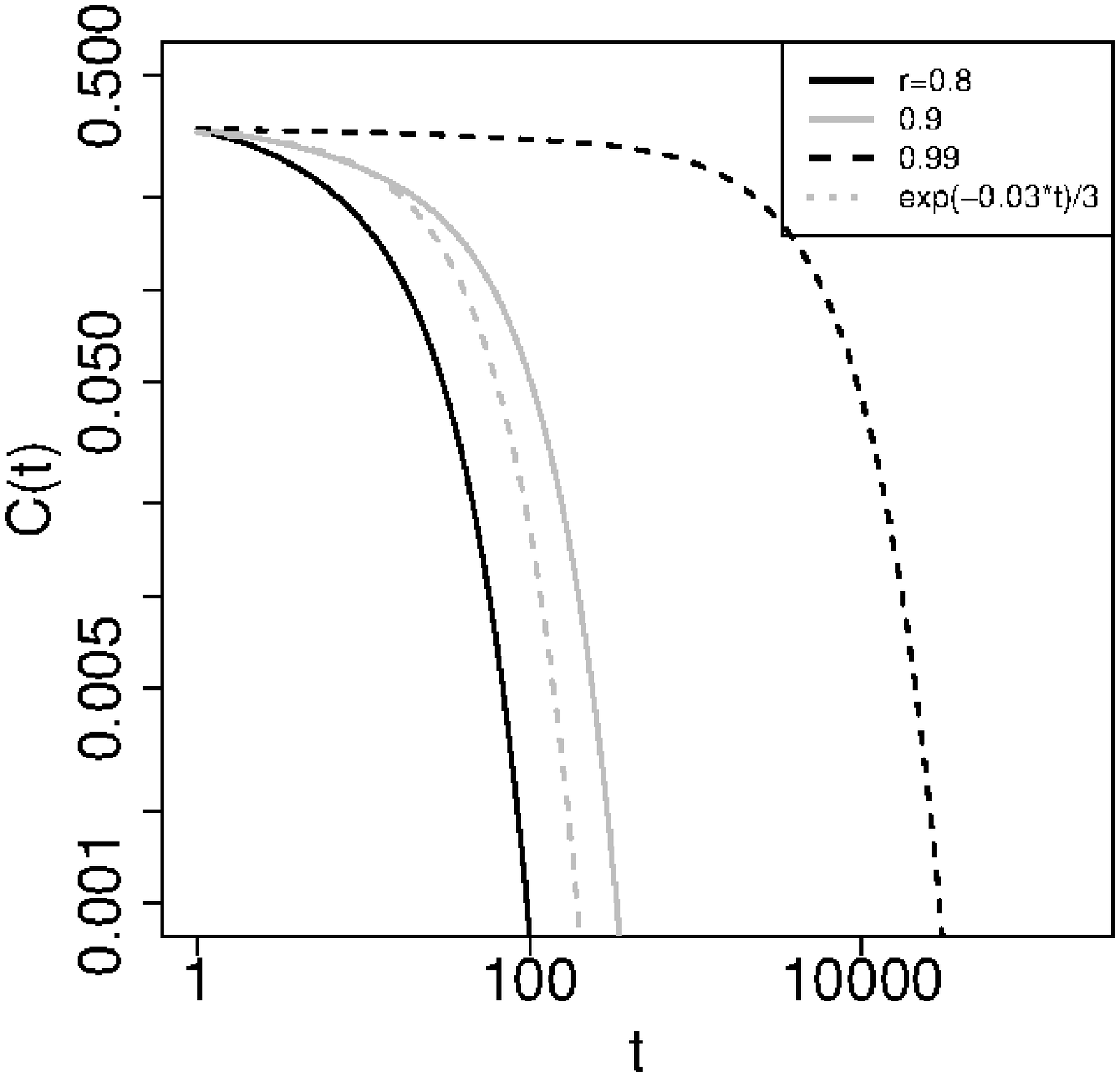}
\hspace{1.6cm} (a)
\end{center}
\end{minipage}
\begin{minipage}{0.5\hsize}
\begin{center}
\includegraphics[clip, width=8cm]{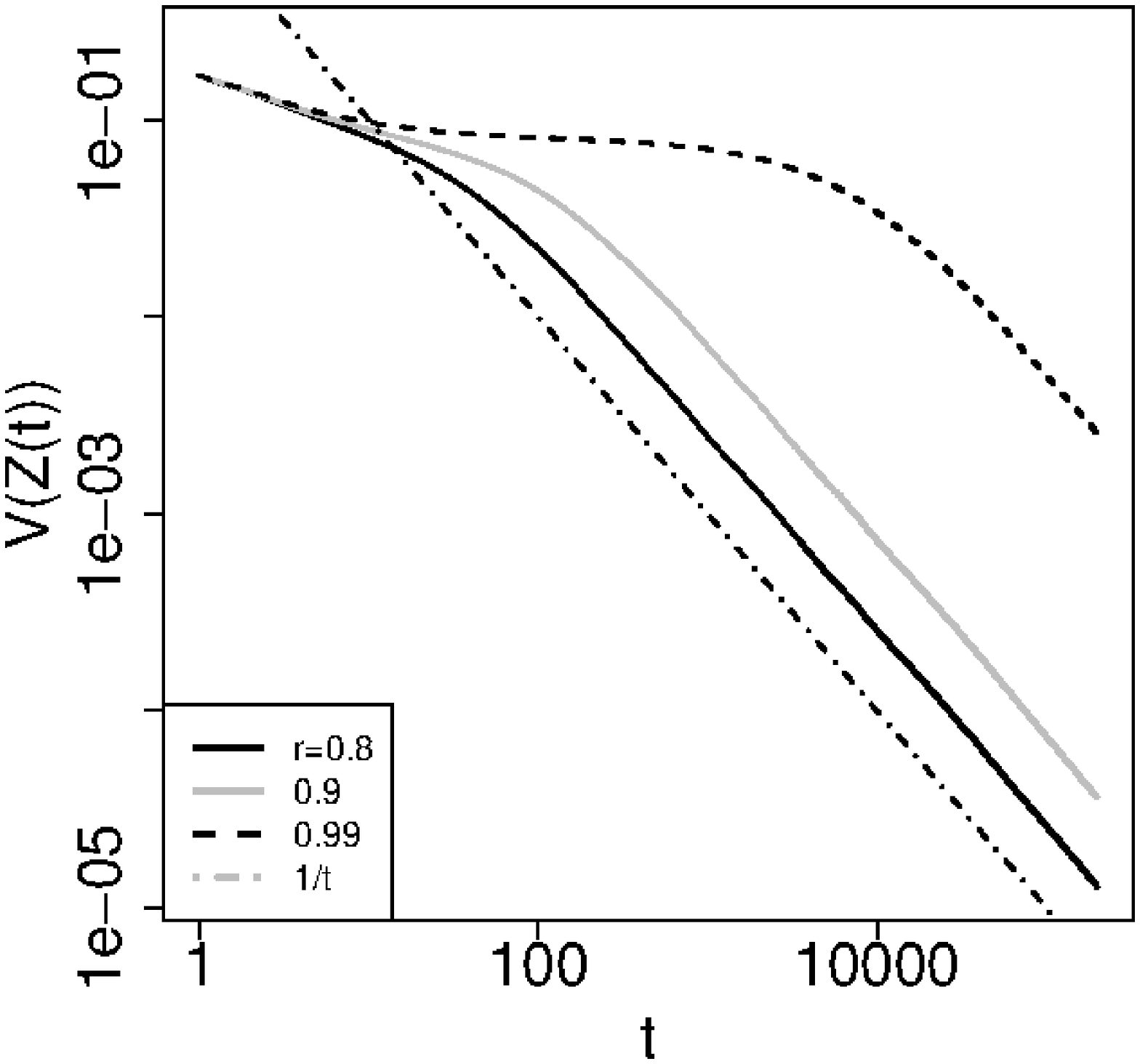}
\hspace{1.6cm} (b)
\end{center}
\end{minipage}
 \end{tabular}
\caption{Plots of (a) $C(t)$ and (b) $V(Z(t))$
  vs. $t$, for $r\in \{0.8,0.9,0.99\}$.
  For comparison, $\exp(-0.03t)/3$ and $1/t$ are potted in (a) and
  in (b), respectively. }
\label{EXP}
\end{center}
\end{figure}

Numerical studies of the system were performed. To estimate
$C(t)$, the recursive relation of Eq. (\ref{eq:self}) is solved for
$t\le 2\times 10^5$. A Monte Carlo
sampling procedure is adopted for the variance of $Z(t)$.
We obtained $10^4$ sample sequences for
$\{X(t)\},t=1,\cdots,2\times 10^5$ and estimated the
variance of $Z(t)$.
Figure \ref{EXP} (a) shows the plot of $C(t)$ vs. $t$.
It is clearly shown that $C(t)$ decays exponentially.
Figure \ref{EXP} (b) shows the plot of V$(Z(t))$ vs. $t$.
For all $r<1\in \{0.8,0.9,0.99\}$, $V(Z(t))$ decays as $1/t$. 
When $r=1$, the $Z(t)$ distribution
converges to the beta distribution.
Hence, there is no phase transition for $r<1$.

\subsubsection{Power-law decay case}

For the case of power-law decay, namely $d_{i}=\frac{1}{(1+i)^{\gamma}}$,
when $\gamma>1$ and $\hat{T}<\infty$, 
the process converges to the delta function.
On the other hand, when $\gamma\leq 1$ and
$\hat{T}$ goes to infinity, the process does not converge.

\begin{figure}[h]
\begin{center}
\begin{tabular}{c}
\begin{minipage}{0.5\hsize}
\begin{center}
\includegraphics[clip, width=8cm]{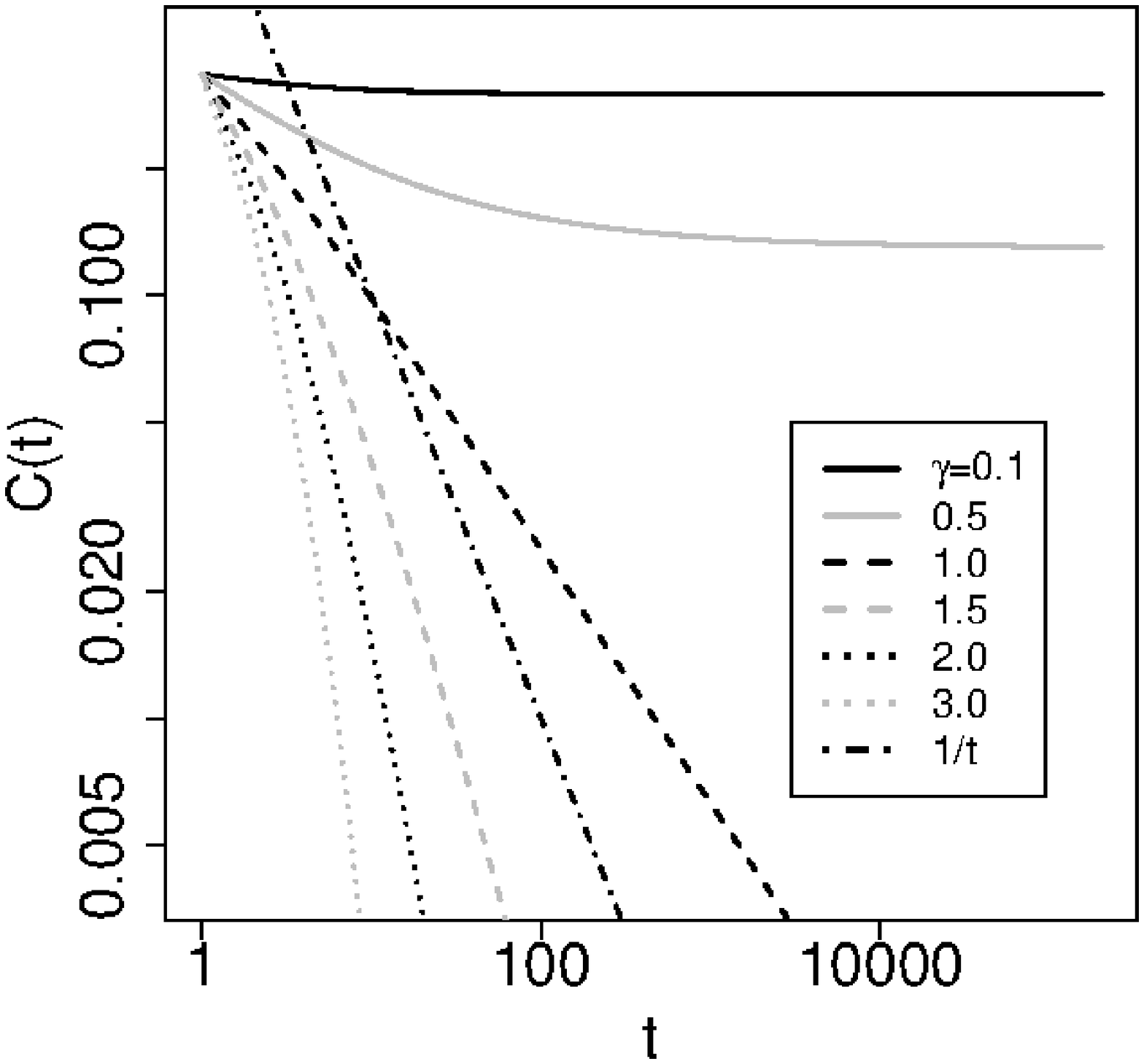}
\hspace{1.6cm} (a)
\end{center}
\end{minipage}
\begin{minipage}{0.5\hsize}
\begin{center}
\includegraphics[clip, width=8cm]{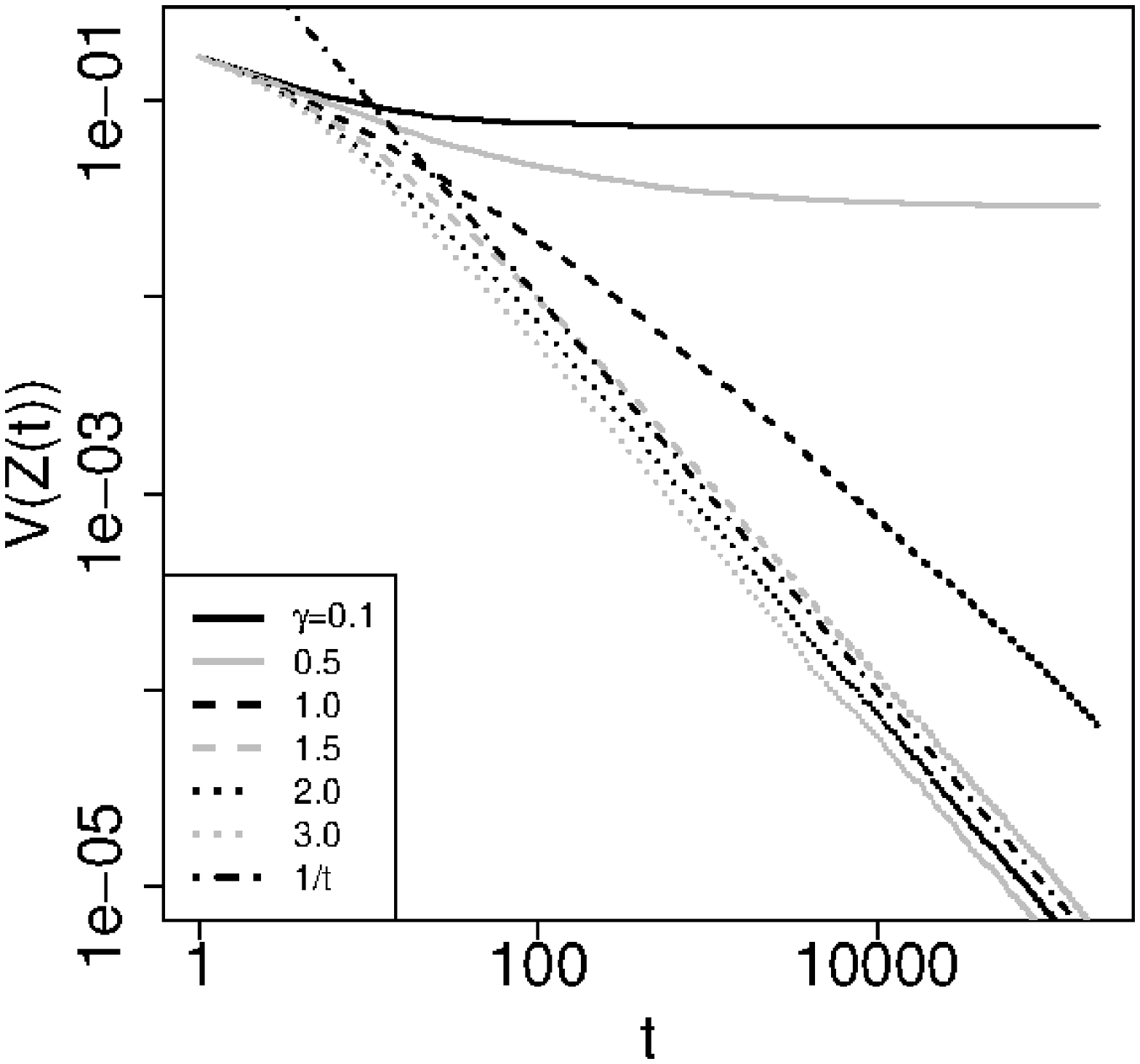}
\hspace{1.6cm} (b)
\end{center}
\end{minipage}
 \end{tabular}
\caption{Plots of (a) $C(t)$ and (b) $V(Z(t))$ vs. $t$, for
  $\gamma \in \{3.0,2.0,1.5,1.0,0.5,0.1\}$.}
\label{POW}
\end{center}
\end{figure}

The behaviors of $C(t)$ and $V(Z(t))$ were investigated by the numerical
method, in the same manner as the exponential decay case.
Fig. \ref{POW} (a) shows the double
logarithmic plot of $C(t)$ vs. $t$. It can be seen that $C(t)$ decays with a power-law form for $\gamma \in \{1.5,2,3\}$.
For small $\gamma$, such as $\gamma = 0.5, 0.1$, the slope is extremely small.
Fig. \ref{POW} (b) shows the double logarithmic plot
of $V(Z(t))$ vs. $t$. For $r=3.0, 2.0, 1.5$, $V(Z(t))$ decays as $1/t$.
At $\gamma=1$, the slope of the decay is less than one.
For $r<1$, the curve is concave down.
These results suggest the validity of the
self-consistent equation analysis.

 To investigate
the phase transition, we apply 
finite-size scaling (FSS) analysis \cite{Mori5}.
We define the relaxation and second-moment correlation times, $\tau(t)$ and $\xi(t)$, respectively, using the $n^{th}$ moment of $C(t)$ as
\begin{eqnarray}
M_{n}(t)&\equiv& \sum_{s=0}^{t-1}C(s)s^n, \nonumber \\
\tau(t)&=& M_{0}(t), \nonumber \\
\xi(t)&=&\sqrt{\frac{M_{2}(t)}{M_{0}(t)}}. \nonumber \\
\label{moment}
\end{eqnarray}
For FSS, we assume that the scaling function, 
$\lim_{t\to \infty}A(st)/A(t)$, for some observable, $A(t)$,
with a scale factor, $s$, is
expressed as a function of $\xi_{t}\equiv \lim_{t\to\infty}\xi(t)/t$ such that
\[
f_{A}(\xi_t)\equiv \lim_{t\to\infty}\frac{A(st)}{A(t)}.
\]

\begin{table}[tbh]
\begin{center}
  \caption{Asymptotic behavior of $C(t)$, and the scaling functions
    $f_{\tau}(\xi_t)$, $f_{\xi}(\xi_t)$, and $\xi_t$.
    The assumed
    asymptotic form of $C(t)$ is given in the second column. The second and the third columns provide
    the scaling functions.
    The last column contains the limit values of $\xi(t)/t$.}
\begin{tabular}{|c|l|c|c|c|} 
\multicolumn{5}{c}{}\\ \hline
No.& Asymptotic behavior & $f_{\tau}(\xi_t)=\lim_{t\to\infty}\frac{\tau(st)}{\tau(t)}$ &$f_{\xi}(\xi_t)=\lim_{t\to\infty}
\frac{\xi(st)}{\xi(t)} $ &  $\xi_t=\lim_{t\to \infty}\xi(t)/t$ \\ \hline \hline
1 &$C(t)\simeq c+\Delta C(t),c>0 $ & $s$& $s$&$1/\sqrt{3}$\\ \hline
2& $C(t)\propto t^{-\delta},0<\delta<1$  & $s^{1-\delta}=s^{\frac{2(\xi/t)^2}{1-(\xi/t)^2}}$ & $s$& ${ \sqrt{\frac{1-\delta}{3-\delta}}}$ \\ \hline 
3&$C(t)\propto t^{-\delta},1<\delta<3$  & $1$ & $s^{(3-\delta)/2}$& $0$ \\ \hline
4&$C(t)\propto t^{-\delta},\delta\ge 3$  & $1$ & $1$& $0$ \\ \hline
\end{tabular}
\end{center}
\label{T1}
\end{table}

We assume the following asymptotic forms for $C(t)$;
\[
C(t) \simeq 
\left\{
\begin{array}{cc}
c+\Delta C(t)  & c>0  \\
c' t^{-\delta}  & c=0
\end{array}
\right.
\]
Here, $c=\lim_{t\to \infty}C(t)$ is the order parameter
of the phase transition and $c'$ is a constant. 
Using the asymptotic forms, we can classify the behavior of
the scaling functions.
We show the results for
$f_{\tau}(\xi_t),f_{\xi}(\xi_t)$ and $\xi_t$
in Table I. (In detail, see Appendix B)

\begin{figure}[h]
\begin{center}
\begin{tabular}{c}
\begin{minipage}{0.5\hsize}
\begin{center}
\includegraphics[clip, width=8cm]{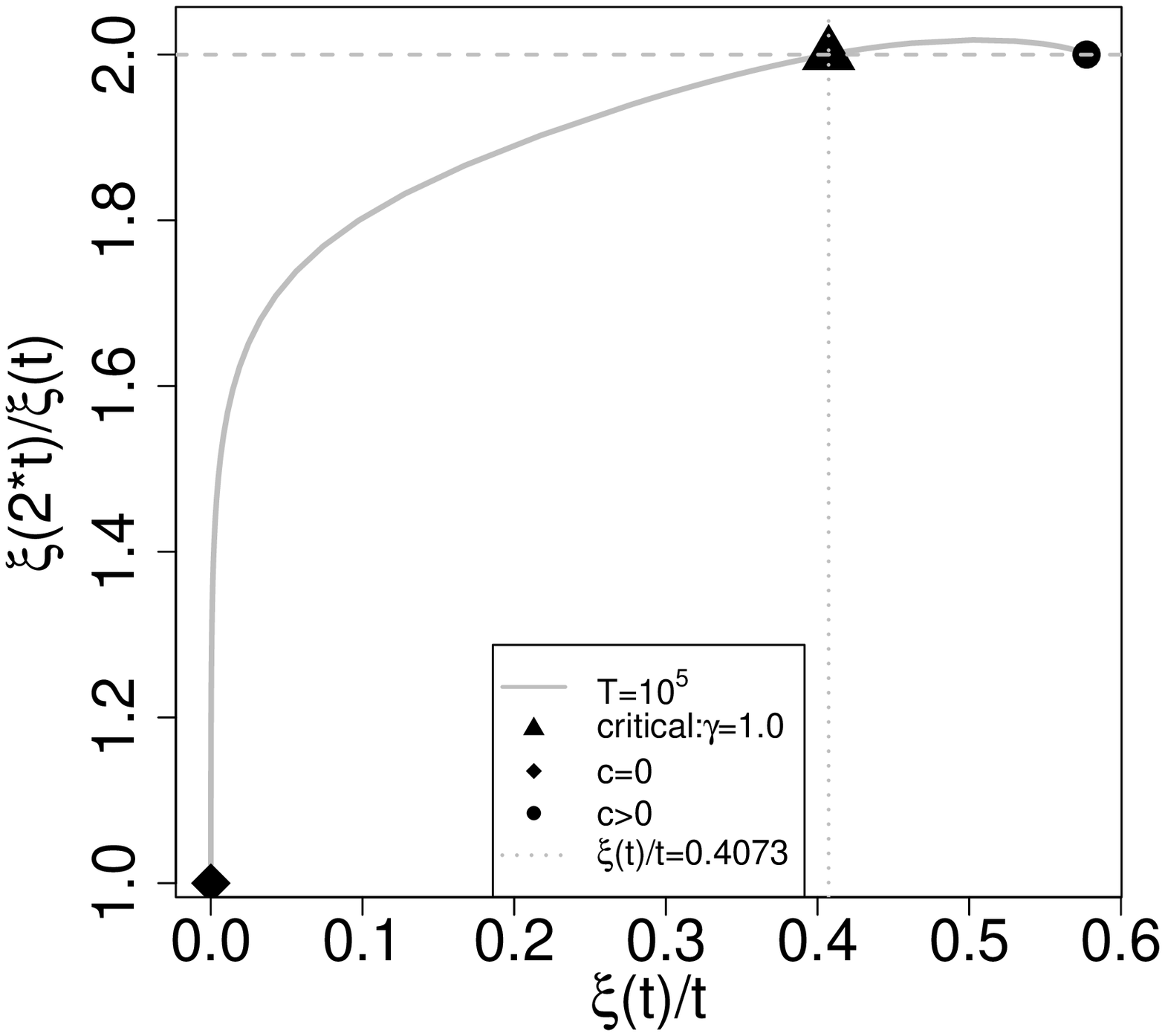}
\hspace{1.6cm} (a)
\end{center}
\end{minipage}
\begin{minipage}{0.5\hsize}
\begin{center}
\includegraphics[clip, width=8cm]{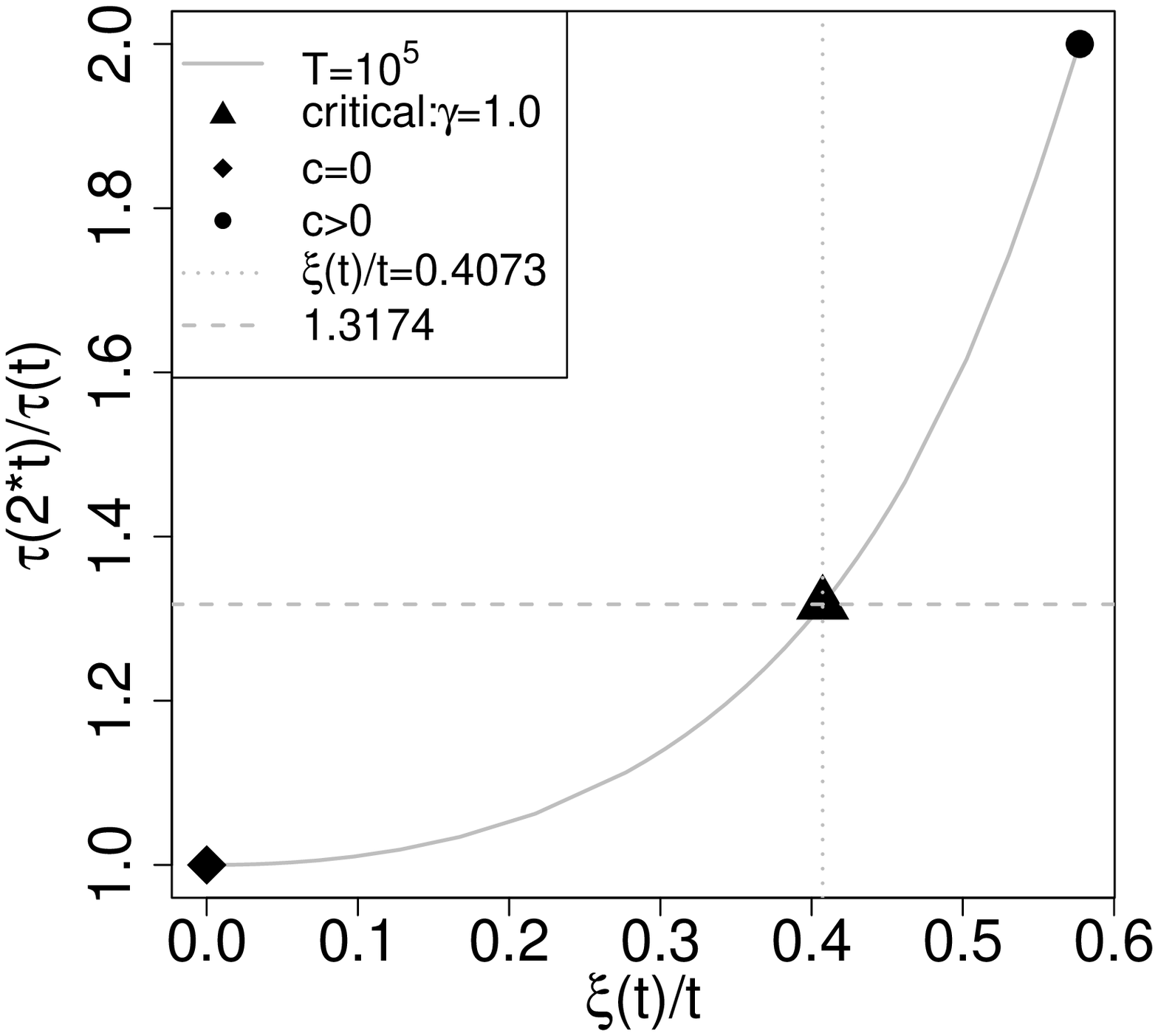}
\hspace{1.6cm} (b)
\end{center}
\end{minipage}
 \end{tabular}
\caption{Plots of (a) $\xi(2t)/\xi(t)$ vs $\xi(t)/t$
  and (b) $\tau(2t)/\tau(t)$ vs
  $\xi(t)/t$.  We adopt $t=10^5$ and
  $a=b=1$. The symbols show the fixed points 
  under the renormalization transformation $t\to 2t$.}
\label{SCALING}
\end{center}
\end{figure}

Figure \ref{SCALING} shows
the numerical estimations of $\xi(2t)/\xi(t)$ and
$\tau(2t)/\tau(t)$ vs. $\xi(t)/t$ with
$t=10^5$.
The symbols show the fixed points under the renormalization transformation
$t\to 2t$. 
There are two stable fixed points at $\xi_t=0$
and $\xi_t=1/\sqrt{3}$, and one unstable fixed point at
$\xi_t=\sqrt{(1-\delta)/(3-\delta)}\simeq 0.4073\equiv \xi_{t}^{c}$.
If $\xi_t>\xi_{t}^{c}$, then $\xi(2t)/\xi(t)>2$ and $\xi(t)/t$ moves
to $1/\sqrt{3}$ under the transformation $t\to 2^{n}t$ and $n\to \infty$.
$\xi(t)$ diverges linearly with the system size, $t$, at the fixed point,
which reflects
the memory of $X(1)$ that retains. 
If $\xi_t<\xi_{t}^{c}$, $\xi(2t)/\xi(t)<2$ and $\xi(t)/t$ moves
to $0$. $\lim_{t\to \infty}\xi(t)<\infty$ and the memory of $X(1)$ is lost
for sufficiently large $t$.
At the stable fixed points of $\xi_t=1/\sqrt{3}$ and at $\xi_t=0$,
$\tau(2t)/\tau(t)$ becomes 2 and 1,
respectively.
From the unstable fixed point at $\xi_t=\xi_{t}^c$, we can estimate $\delta$ using 
$f_{\tau}(\xi_t^c)=2^{1-\delta}\simeq 1.3174$. This estimation is in accordance with
the estimation from  $\xi_{t}^{c}=\sqrt{(1-\delta)/(3-\delta)}\simeq 0.4073$.
These results support the phase transition between the two phases, 
$C(t)\simeq c+\Delta C(t),c>0$ and  $C(t)\propto t^{-\delta},\delta>1$,
in the limit $t\rightarrow \infty$. At the critical point $\gamma=1$, $\xi_t=\xi_t^c$ and
$C(t)\propto t^{-\delta}$ with $0<\delta<1$.

\begin{figure}[h]
\begin{center}
\begin{tabular}{c}
\begin{minipage}{0.33\hsize}
\begin{center}
\includegraphics[clip, width=5cm]{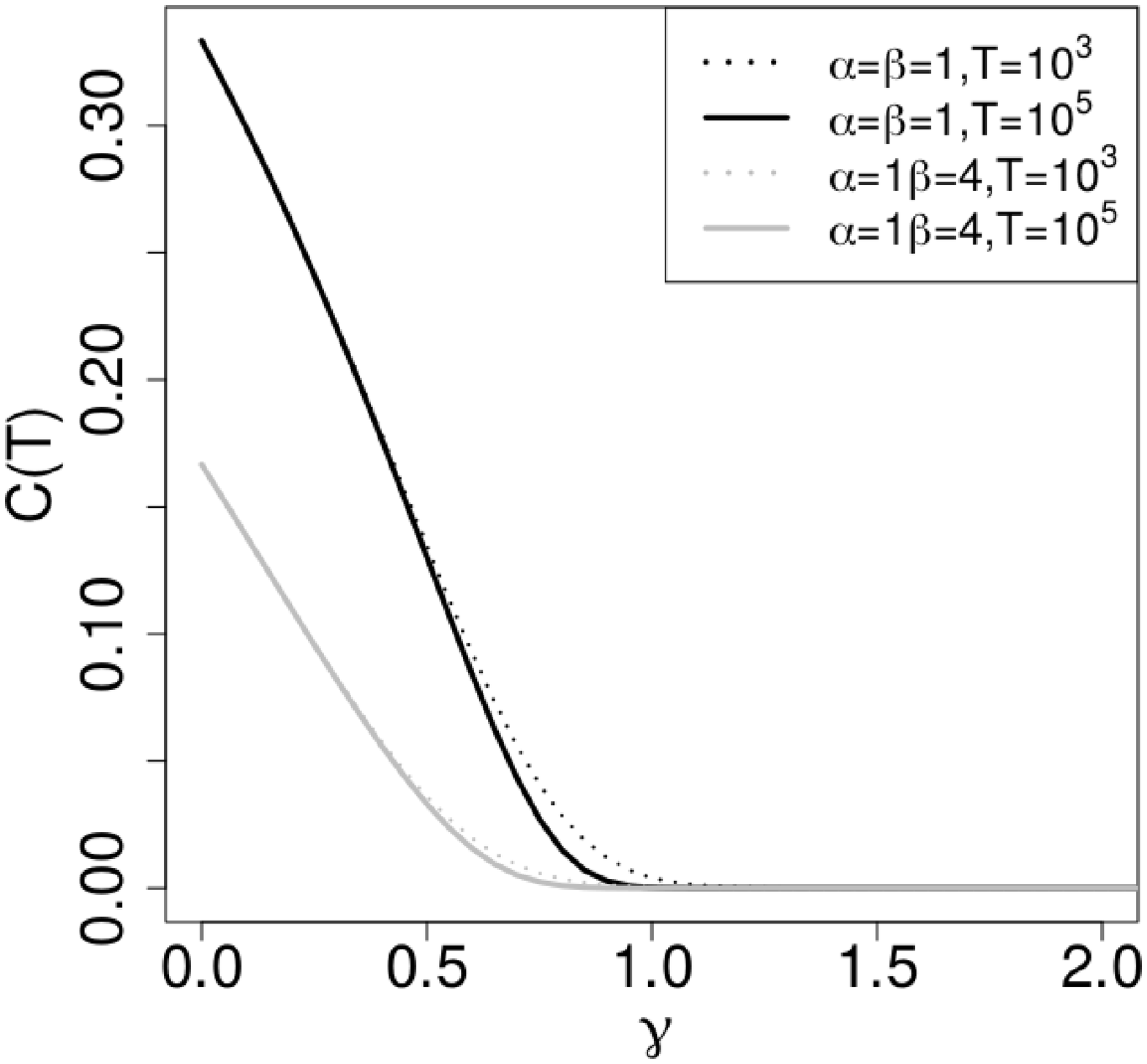}
\hspace{1.6cm} (a)
\end{center}
\end{minipage}
\begin{minipage}{0.33\hsize}
\begin{center}
\includegraphics[clip, width=5cm]{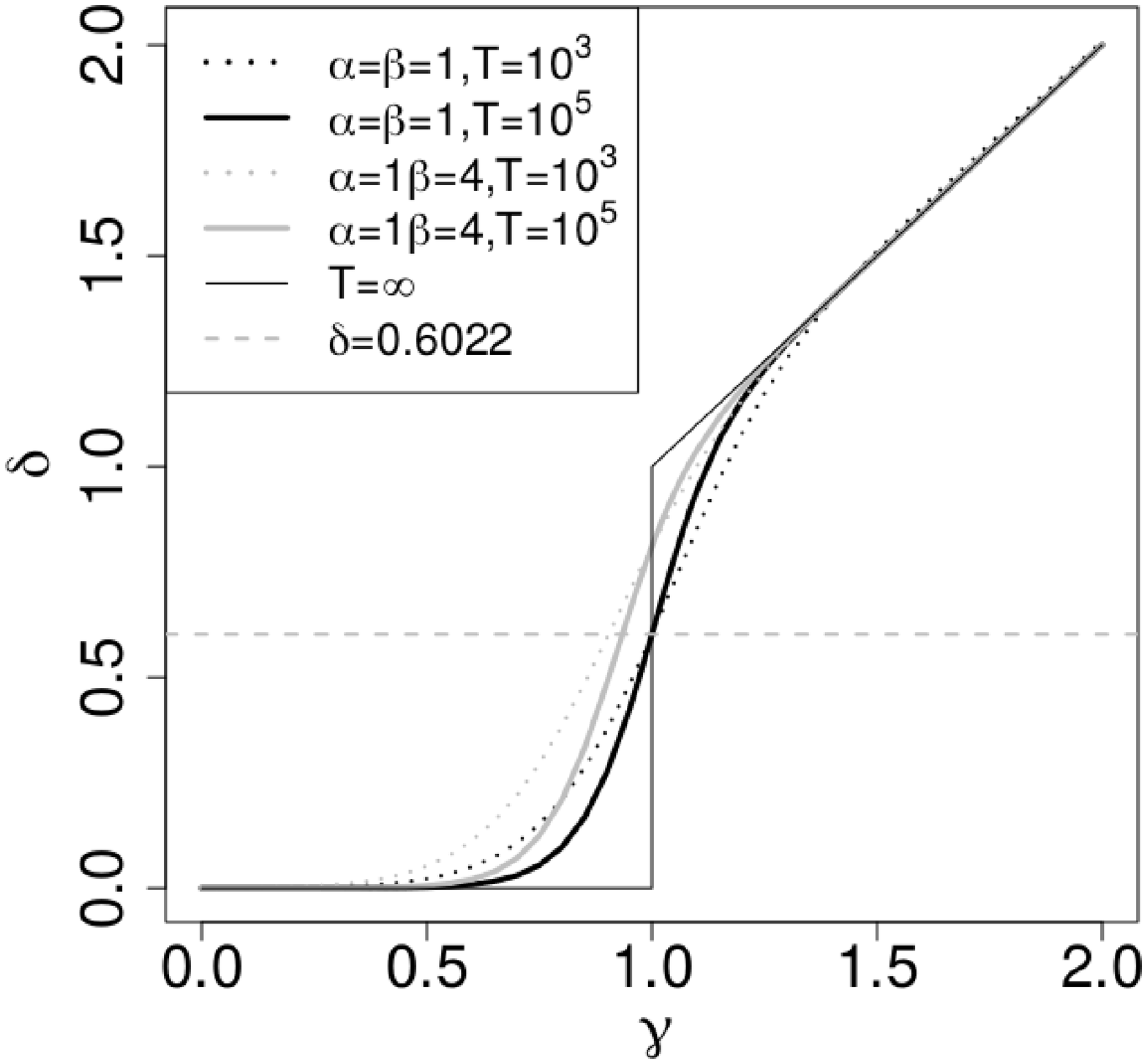}
\hspace{1.6cm} (b)
\end{center}
\end{minipage}
\begin{minipage}{0.33\hsize}
\begin{center}
\includegraphics[clip, width=5cm]{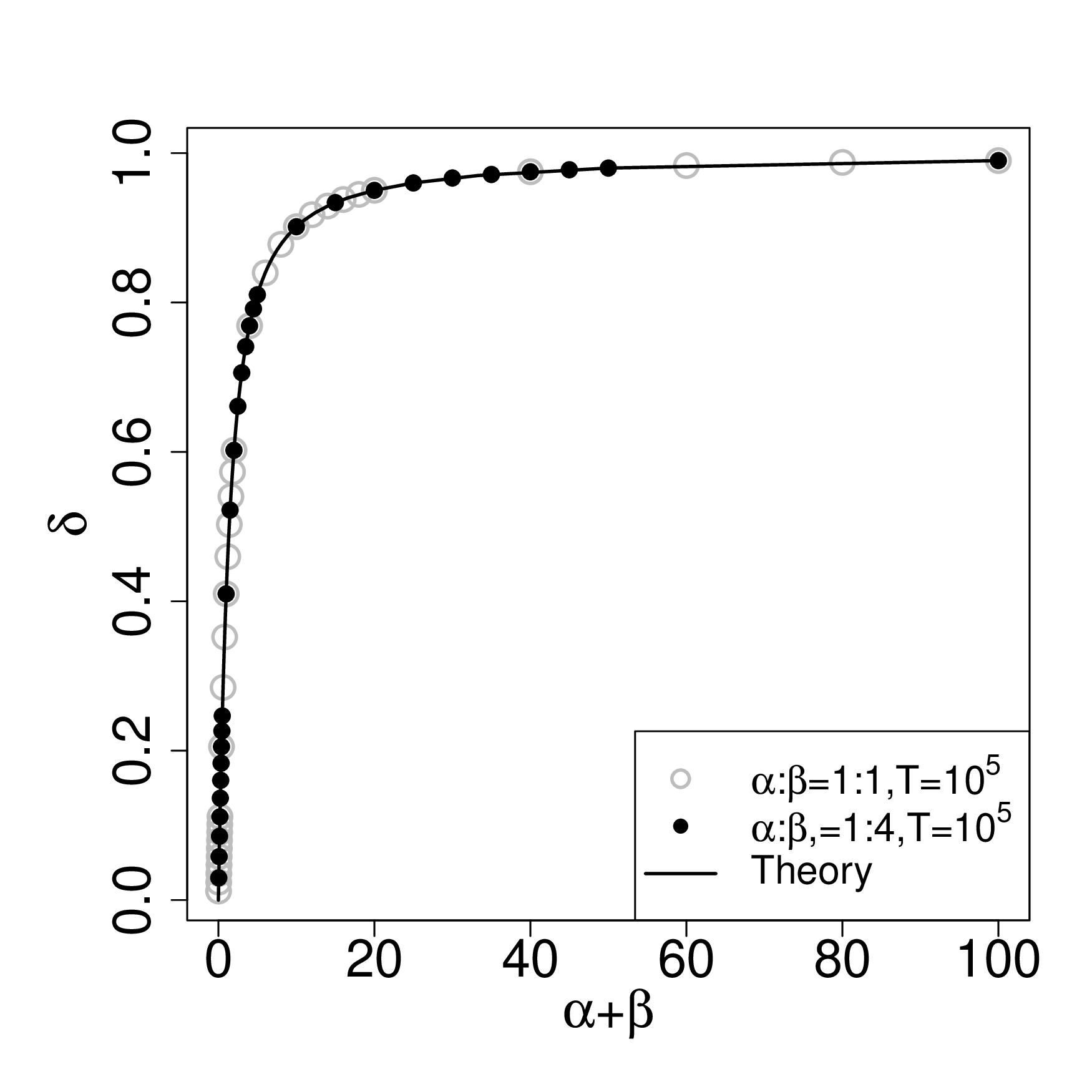}
\hspace{1.6cm} (c)
\end{center}
\end{minipage}

 \end{tabular}
\caption{Plots of (a) $C(t)$ and (b) $\delta$ vs. $\gamma$.
  We adopt $t=2\times 10^3,2\times 10^5$, and $(\alpha,\beta)=(1,1)$ and $(1,4)$.
   The conjecture presented
  in the main text is plotted in (b) with the thin solid line.
  $\delta$ for $\gamma=1$ and $\alpha=\beta=1$ (thick solid and dotted black lines, respectively)
  are estimated by $\xi_{t}^{c}=\sqrt{(1-\delta)/(3-\delta)}$ and 
  $\xi_{t}^{c}$ in Figure \ref{SCALING}.
  (c) Plot of $\delta$ at $\gamma=1$ vs. $\alpha+\beta$.
  We set the ratios $\alpha:\beta=1:1$ and $1:4$, and change $\alpha+\beta$.
  The solid line shows the $\delta$ estimation by solving Eq. (\ref{eq:delta_c}).
}
\label{c_and_delta}
\end{center}
\end{figure}

We estimate $c$ by $C(2t)$  and 
 $\delta$ by
$\log_{2}C(t)/C(2t)$ with $t=10^3$ and $10^5$.
By comparing the values for $t=10^3$ and $10^5$,
one can anticipate the limit behavior $t\to \infty$.
The results are shown in Fig. \ref{c_and_delta}.
Figure \ref{c_and_delta} (a) shows $C(2t)$ vs. $\gamma$.
For $\gamma>1$, $C(2t)$ is almost zero.
For $\gamma<1$, $C(2t)$ is positive. The derivative of $c$ at
$\gamma=1$ is seemingly continuous.
Figure \ref{c_and_delta} (b) shows $\delta$ vs $\gamma$ with
$(\alpha,\beta)=(1,1)$ and $(1,4)$.
For $\gamma>1$, one can anticipate that $\delta=\gamma$
by observing the change from $t=10^3$ to $10^5$.
For $\gamma<1$, $\delta=0$ which suggests that $c>0$.
At the critical point $\gamma=1$, $\delta$ depends on $(\alpha,\beta)$.

Next, we investigated $\delta$ at the critical point $\gamma=1$.
We assume that $C(t)\propto t^{-\delta}$.
Eq. (\ref{eq:self})  can be approximated in the continuous limit as
\begin{equation}
  C(t)=t^{-\delta}=\frac{\int^{t} (s-1)^{-\delta} d(t-s)ds}
  {\alpha+\beta+\int^{t}d(t-s)ds}.
\end{equation}
By the following change of variables, $(t+1)\mu=s$, we obtain
\begin{equation}
  \alpha+\beta\simeq\int_{1/(t+1)}^{t/(t+1)}\mu^{-\delta}(1-\mu)^{-1}d\mu-\ln t.
  \label{eq:delta_c}. 
\end{equation}
We see that $\delta$ depends on $\alpha$ and $\beta$ through the combination $\alpha+\beta$.
In the limit $t \rightarrow \infty$, when $\delta=1$ and $0$,
$\alpha+\beta=0$ and $\alpha+\beta\rightarrow \infty$, respectively.
The critical exponent $\delta$ is in the range $\delta \in (0,1)$.
Figure \ref{c_and_delta} (c) shows $\delta$ vs. $\alpha+\beta$ for $\gamma=1$.
We adopt two cases $\alpha:\beta=1:1$ and $1:4$.
The symbols show the results of the numerical
estimation, and the solid line shows the results by numerically
solving Eq. (\ref{eq:delta_c}). The results for $\alpha:\beta=1:1$ and $1:4$ 
collapses onto the same curve vs. $\alpha+\beta$,
which confirms that $\delta$ depends on
$\alpha$ and $\beta$ through $\alpha+\beta$.

\section{Is the temporal correlation decay exponential or power?}
In this section, we use three data sets from the default data. Two sets are rating agency data, and the other is from a Japanese company.

\subsection{Standard \& Poor's data}
\begin{figure}[h]
\begin{center}
\begin{tabular}{c}
\begin{minipage}{0.5\hsize}
\begin{center}
\includegraphics[clip, width=8cm]{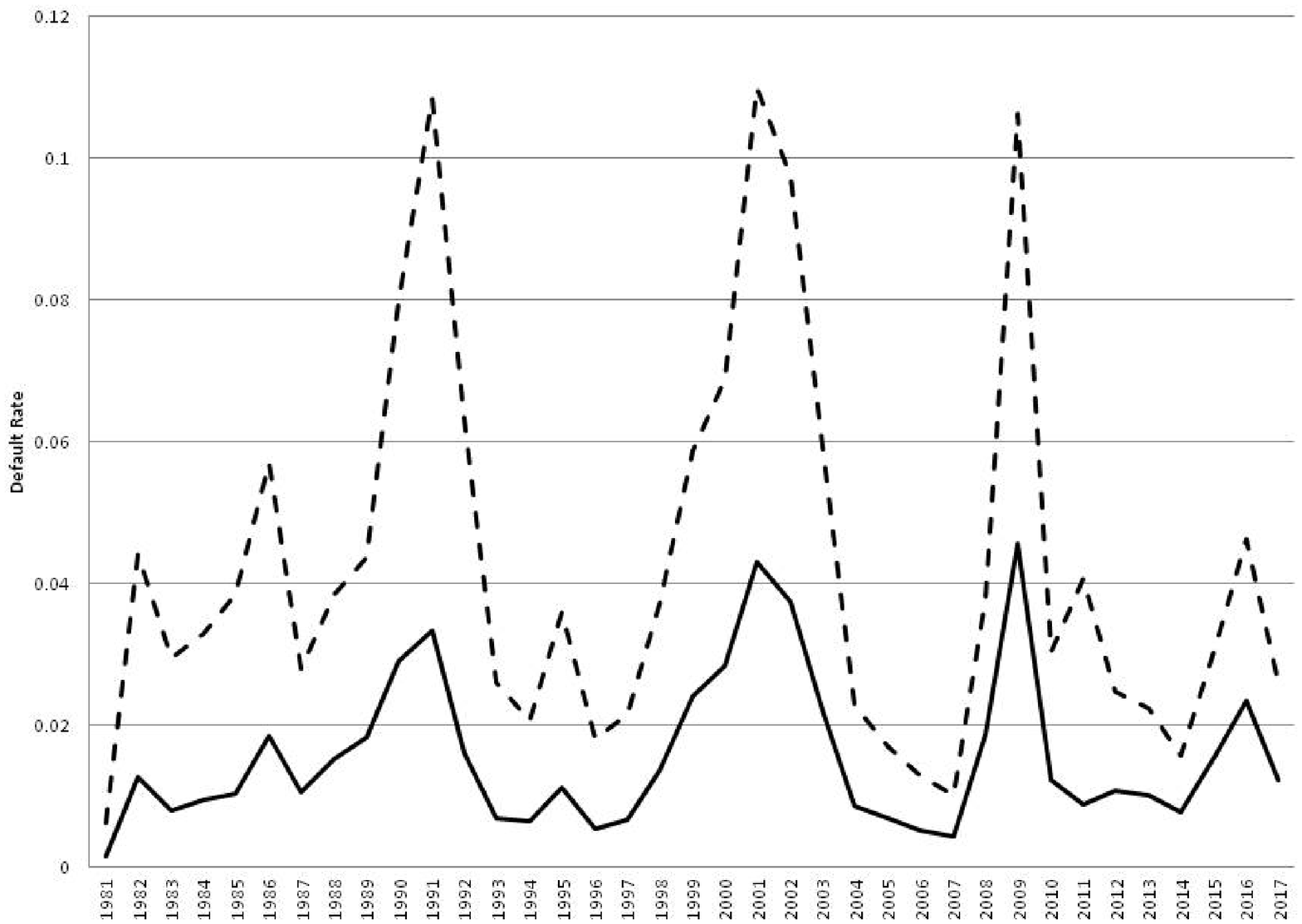}
\hspace{1.6cm} (a)
\end{center}
\end{minipage}
\begin{minipage}{0.5\hsize}
\begin{center}
\includegraphics[clip, width=8cm]{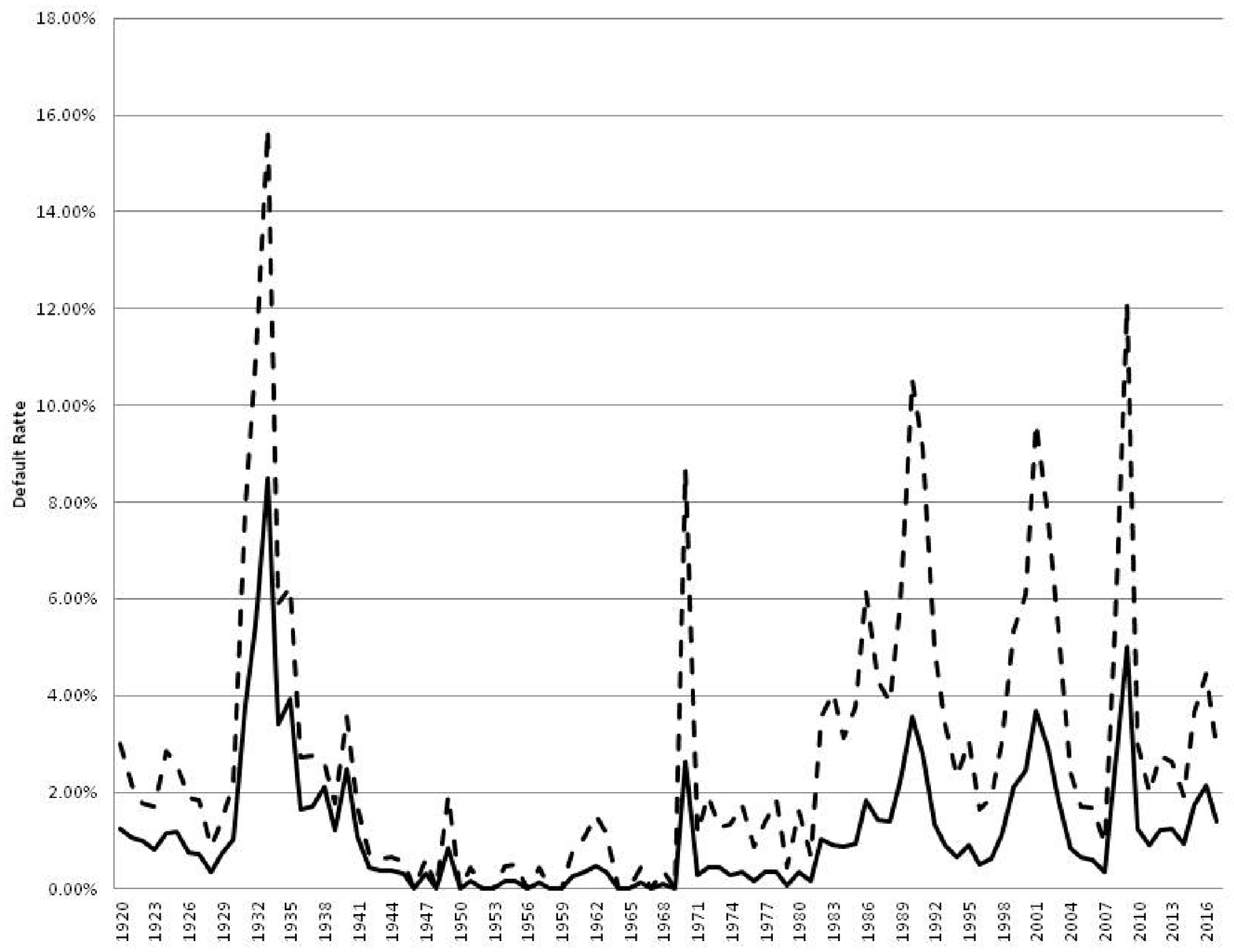}
\hspace{1.6cm} (b)
\end{center}
\end{minipage}
 \end{tabular}
\caption{ (a): S\&P Default Rate from 1981-2017. (b)Moody's Default Rate from 1920-2017. The solid and dotted lines correspond to all the samples and the speculative grade, respectively, below BBB+(Baa3).}
\label{3}
\end{center}
\end{figure}

As discussed in the previous section, temporal correlation is a critical issue for determining whether there is an exponential or a power decay.
This affects whether the parameters are estimated correctly.
 In this section we investigate the temporal correlation using empirical data.
First, the S\&P default data from 1981 to 2017 \cite{Data1} are used. The average PD is 1.58 $\%$ for all ratings and 3.09 $\%$ for speculative ratings. 
A speculative grade rating represents the rating under BBB-(Baa3).
In Fig. \ref{3} (a) we show the historical default rate.
The solid and dotted lines correspond to all the samples and the speculative grade, respectively, below BBB+(Baa3).

\begin{figure}[h]
\begin{center}
\begin{tabular}{c}
\begin{minipage}{0.5\hsize}
\begin{center}
\includegraphics[clip, width=8cm]{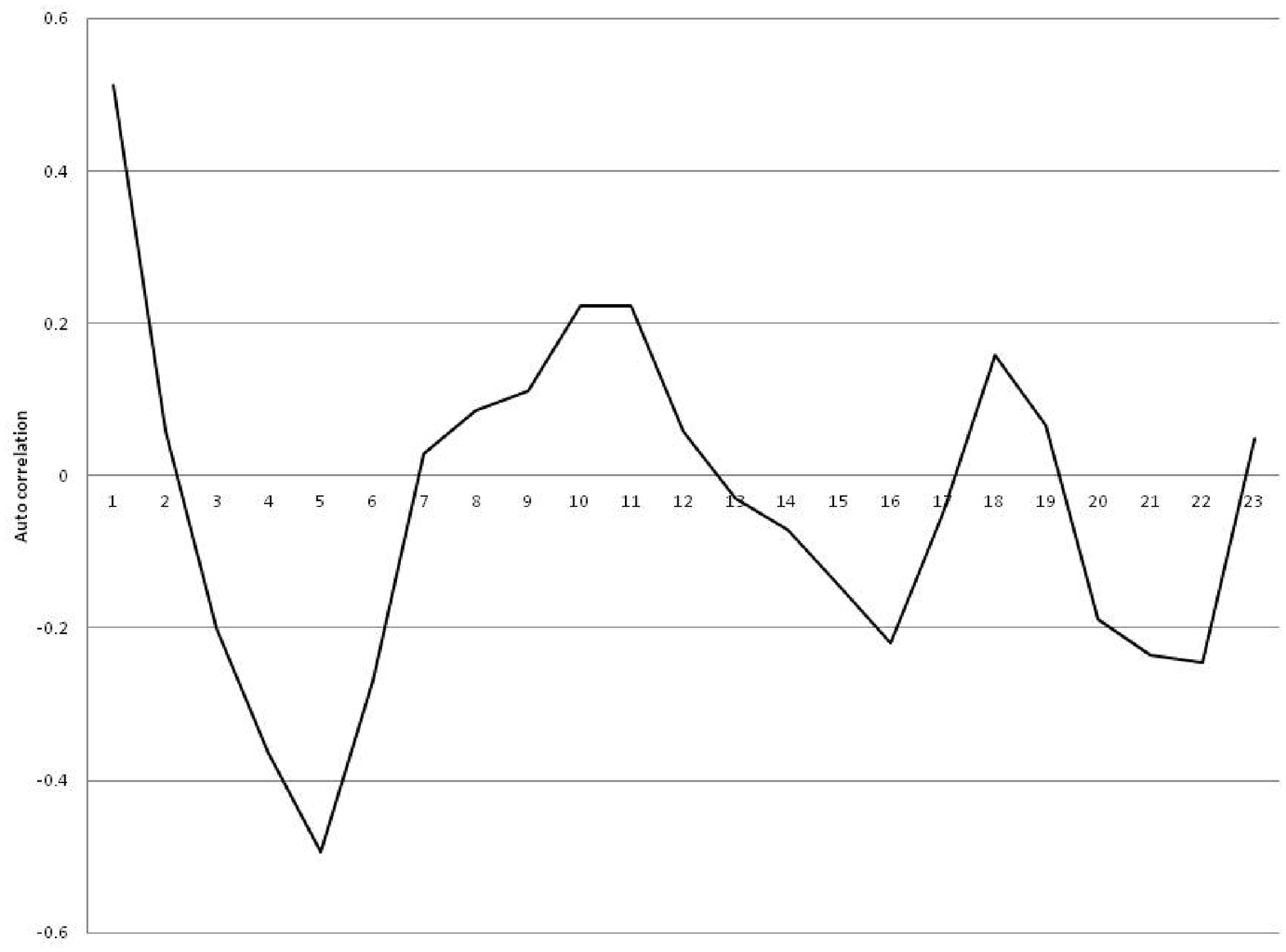}
\hspace{1.6cm} (a)
\end{center}
\end{minipage}
\begin{minipage}{0.5\hsize}
\begin{center}
\includegraphics[clip, width=8cm]{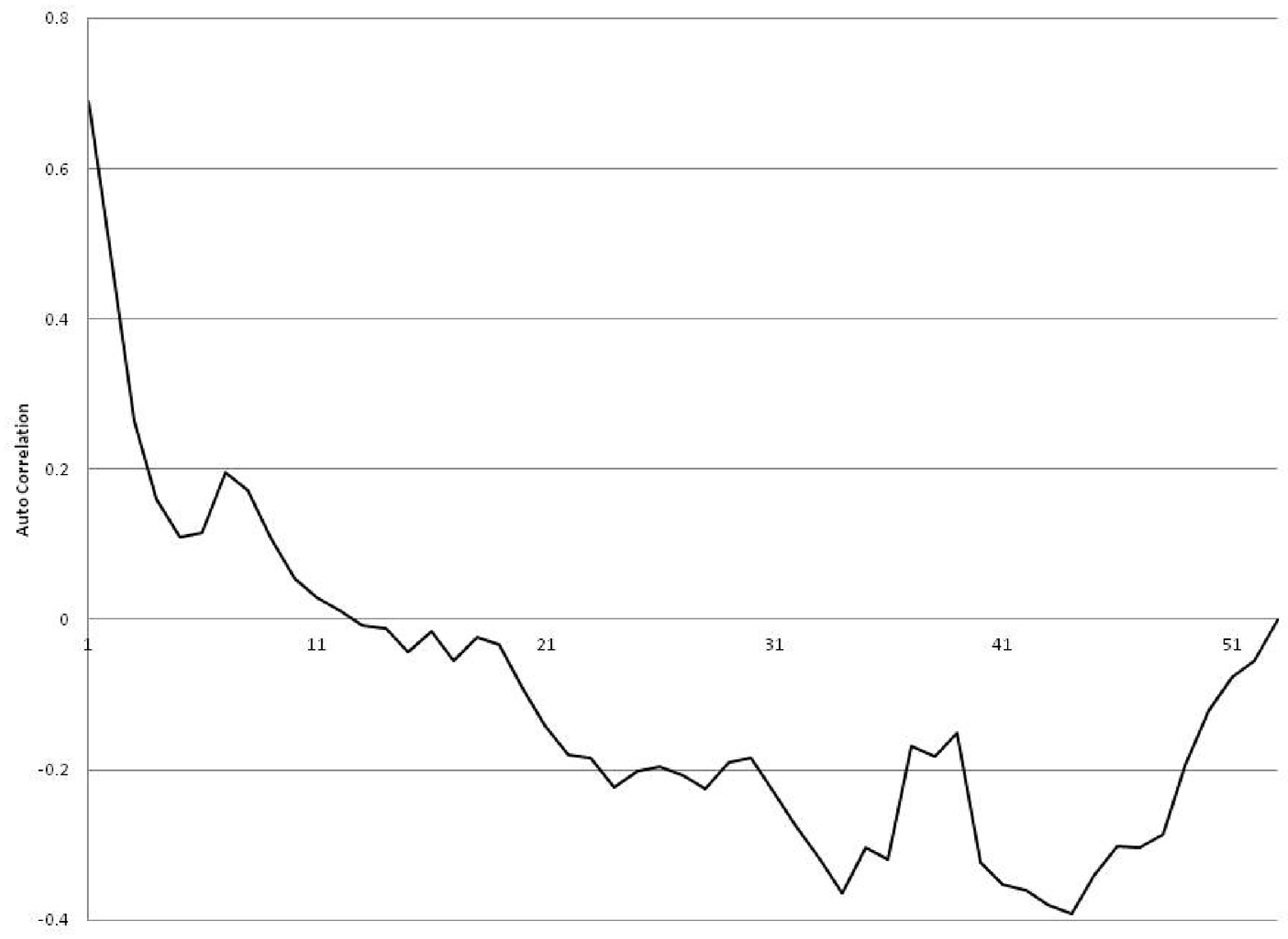}
\hspace{1.6cm} (b)
\end{center}
\end{minipage}
 \end{tabular}
\caption{ (a) S\&P autocorrelation of the default rate from 1981-2017. (b) Moody's autocorrelation of the default rate from 1920-2017. }
\label{4}
\end{center}
\end{figure}

\begin{figure}[h]
\begin{center}
\begin{tabular}{c}
\begin{minipage}{0.5\hsize}
\begin{center}
\includegraphics[clip, width=8cm]{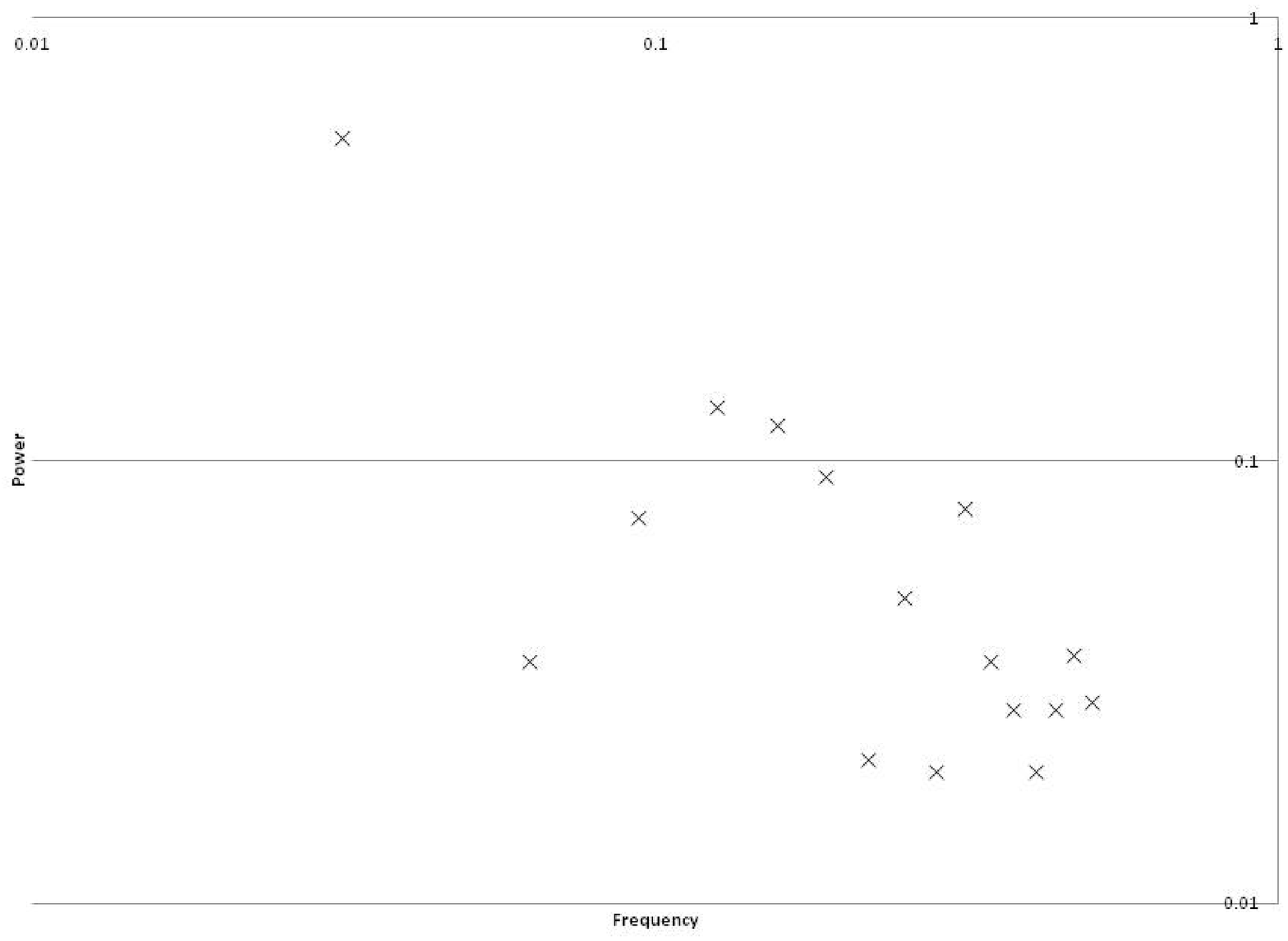}
\hspace{1.6cm} (a)
\end{center}
\end{minipage}
\begin{minipage}{0.5\hsize}
\begin{center}
\includegraphics[clip, width=8cm]{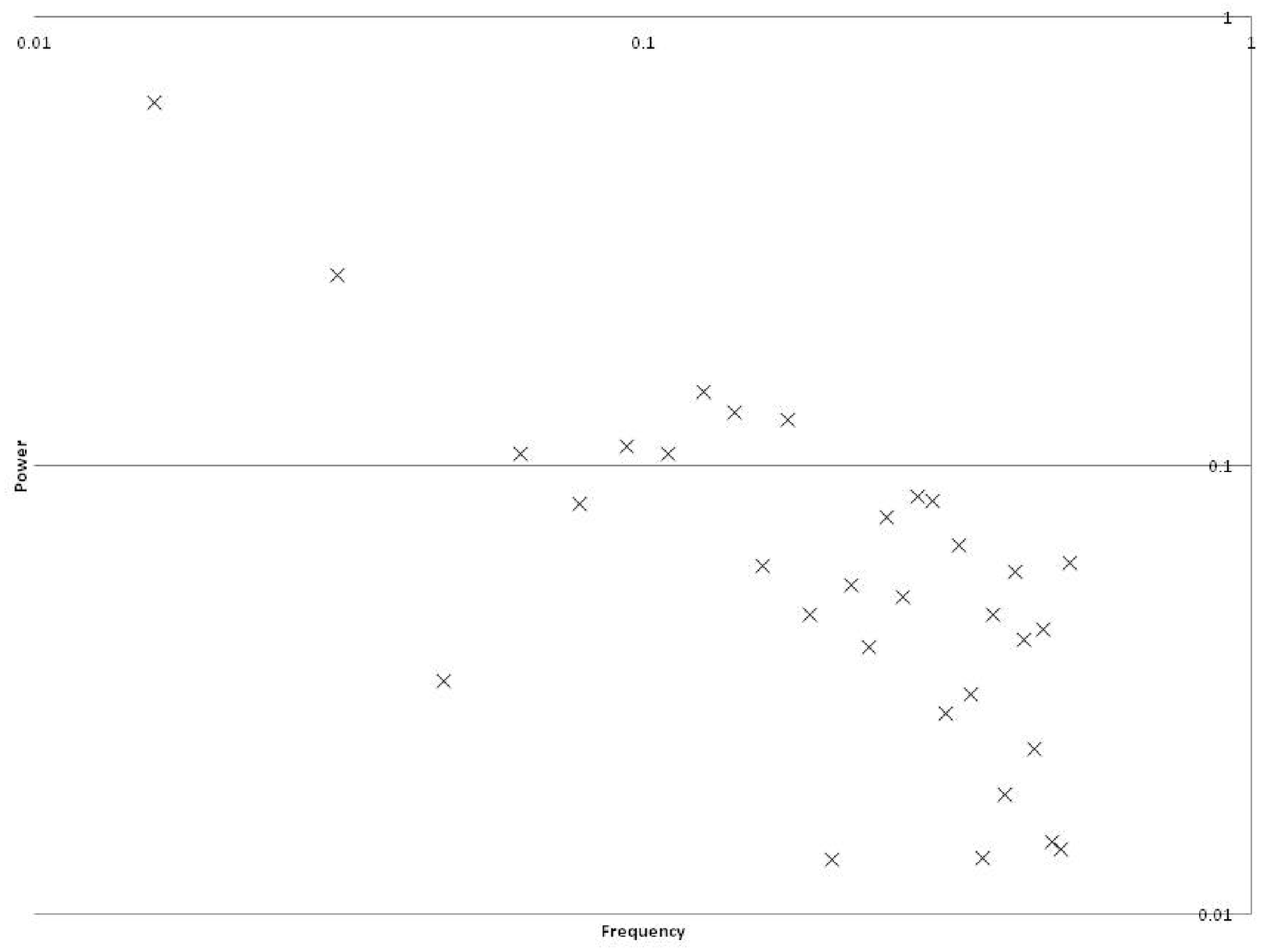}
\hspace{1.6cm} (b)
\end{center}
\end{minipage}
 \end{tabular}
\caption{ (a)Power spectrum for S\&P Default Rate from 1981-2017. (b) Power spectrum for Moody's Default Rate from 1920-2017.}
\label{5}
\end{center}
\end{figure}

The autocorrelation is shown in Fig. \ref{4} (a). The x-axis represents the year. The exponential decay and cyclical increase are confirmed. This represents the cyclical bubbles and their collapse in recent years. However, it is difficult to confirm whether the decay is exponential or power-law from the autocorrelation data alone.
Therefore, a Fourier transformation was applied to the PD data in Fig. \ref{5} (a), but it was still difficult to obtain confirmation because the data was annual, and its size was not very large.  

\subsection{Moody's data}

Next, we used Moody's default data from 1920 to 2017 for 98 years \cite{Data2}. It includes the Great Depression in 1929 and Great Recession in 2008.
It is one of the longest sets of default data \cite{Hec}.
The average default rate is 1.56$\%$ for all the ratings and $3.87\%$
 for the speculative ratings. In Fig. \ref{3} (b), we show the historical default rate.


The autocorrelation is shown in Fig. \ref{4} (b). The x-axis represents the year. The exponential decay is confirmed for a short time. Over the long historical data, we cannot confirm the cyclical trend that was observed in recent years. We applied a Fourier transformation to the default ratio data in Fig. \ref{5} (b). as it is difficult to confirm whether the decay is exponential or power-law from the autocorrelation alone.


\subsection{Risk Data Bank data}

\begin{figure}[h]
\begin{center}
\begin{tabular}{c}
\begin{minipage}{0.5\hsize}
\begin{center}
\includegraphics[clip, width=8cm]{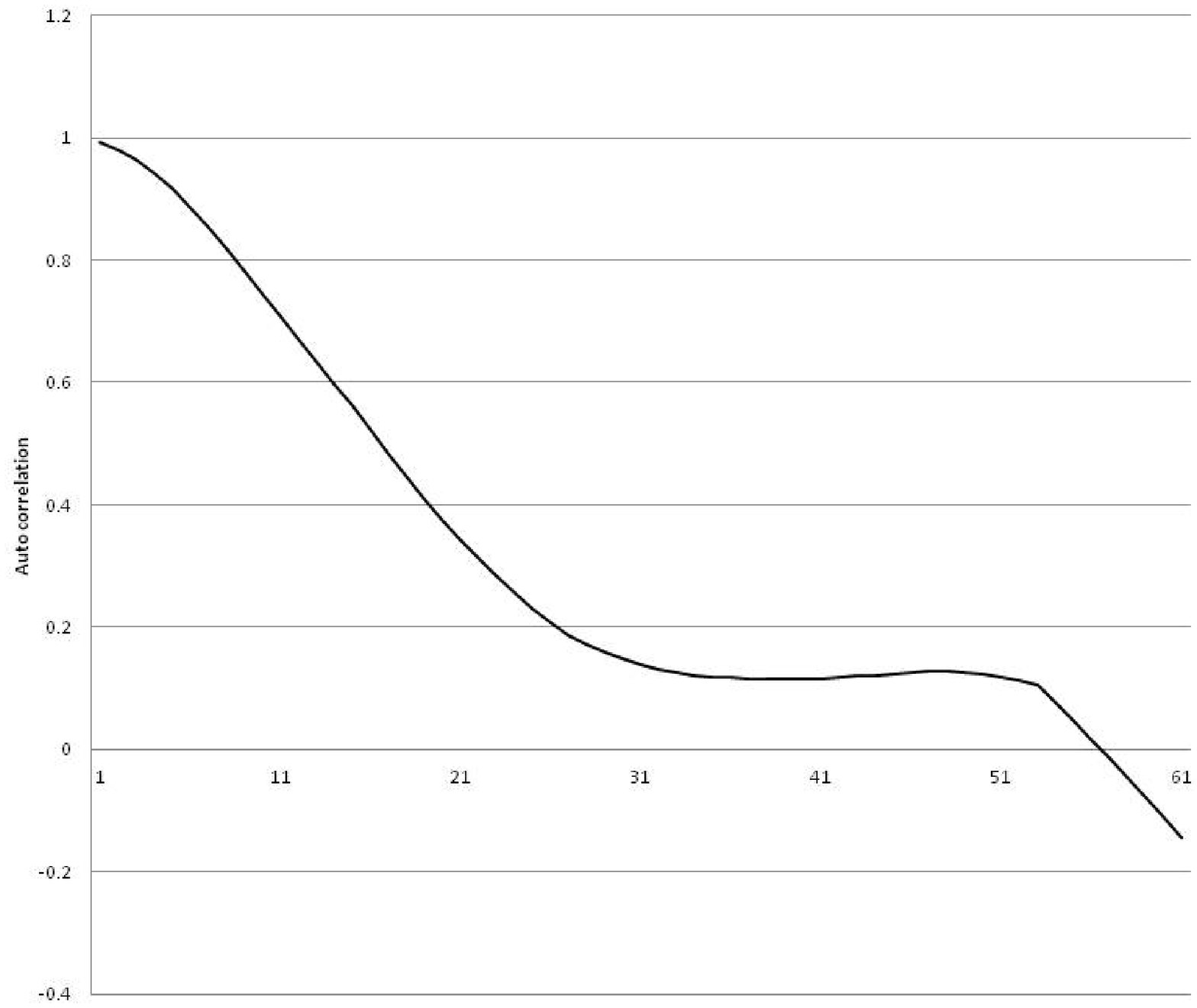}
\hspace{1.6cm} (a)
\end{center}
\end{minipage}
\begin{minipage}{0.5\hsize}
\begin{center}
\includegraphics[clip, width=8cm]{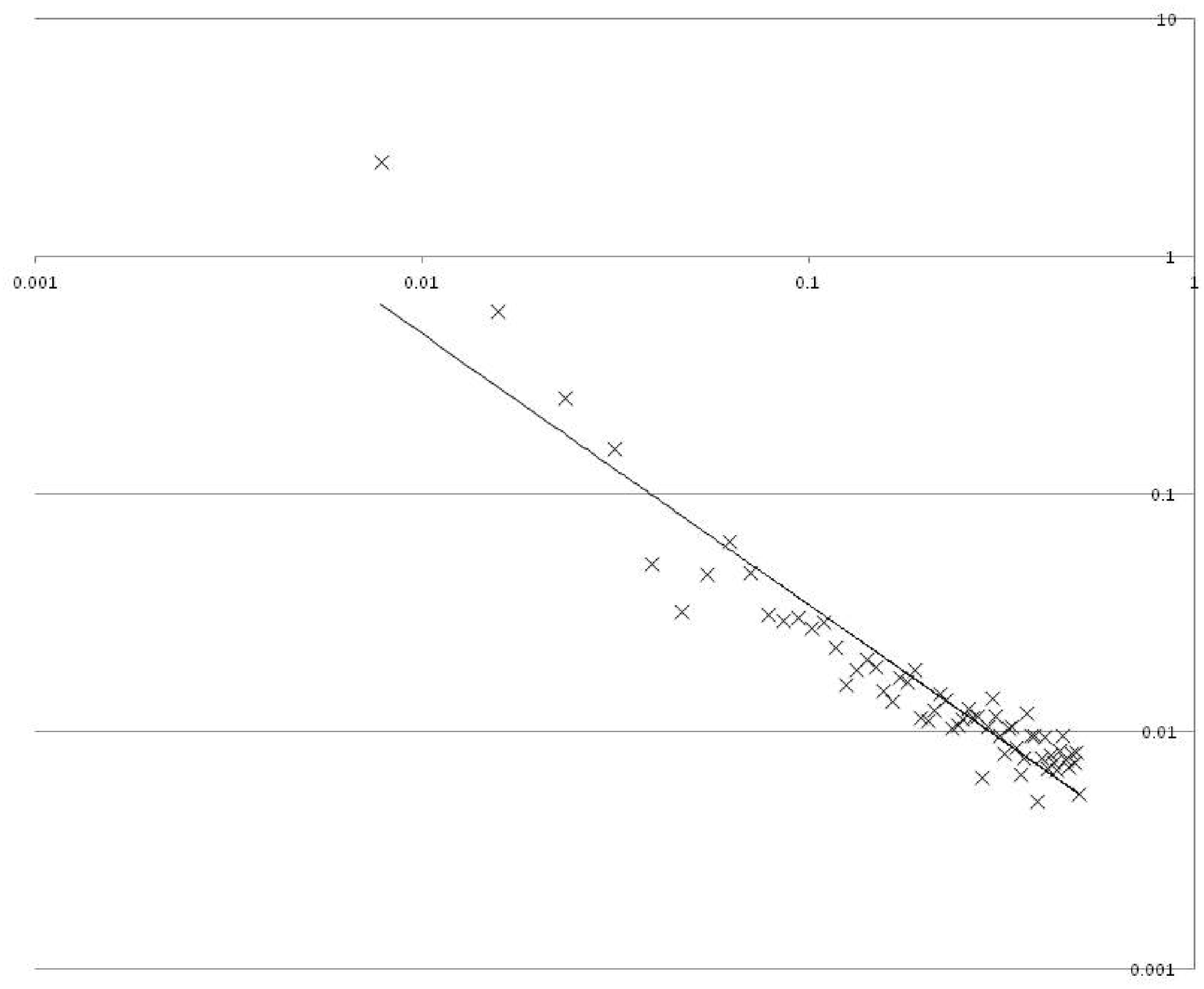}
\hspace{1.6cm} (b)
\end{center}
\end{minipage}
 \end{tabular}
\caption{ (a) Risk data bank autocorrelation of the default rate. (b) Risk data bank power spectrum for the default rate.}
\label{6}
\end{center}
\end{figure}

\begin{figure}[htbp]
\begin{tabular}{ccc}    
(a)\includegraphics[width=4.5cm]{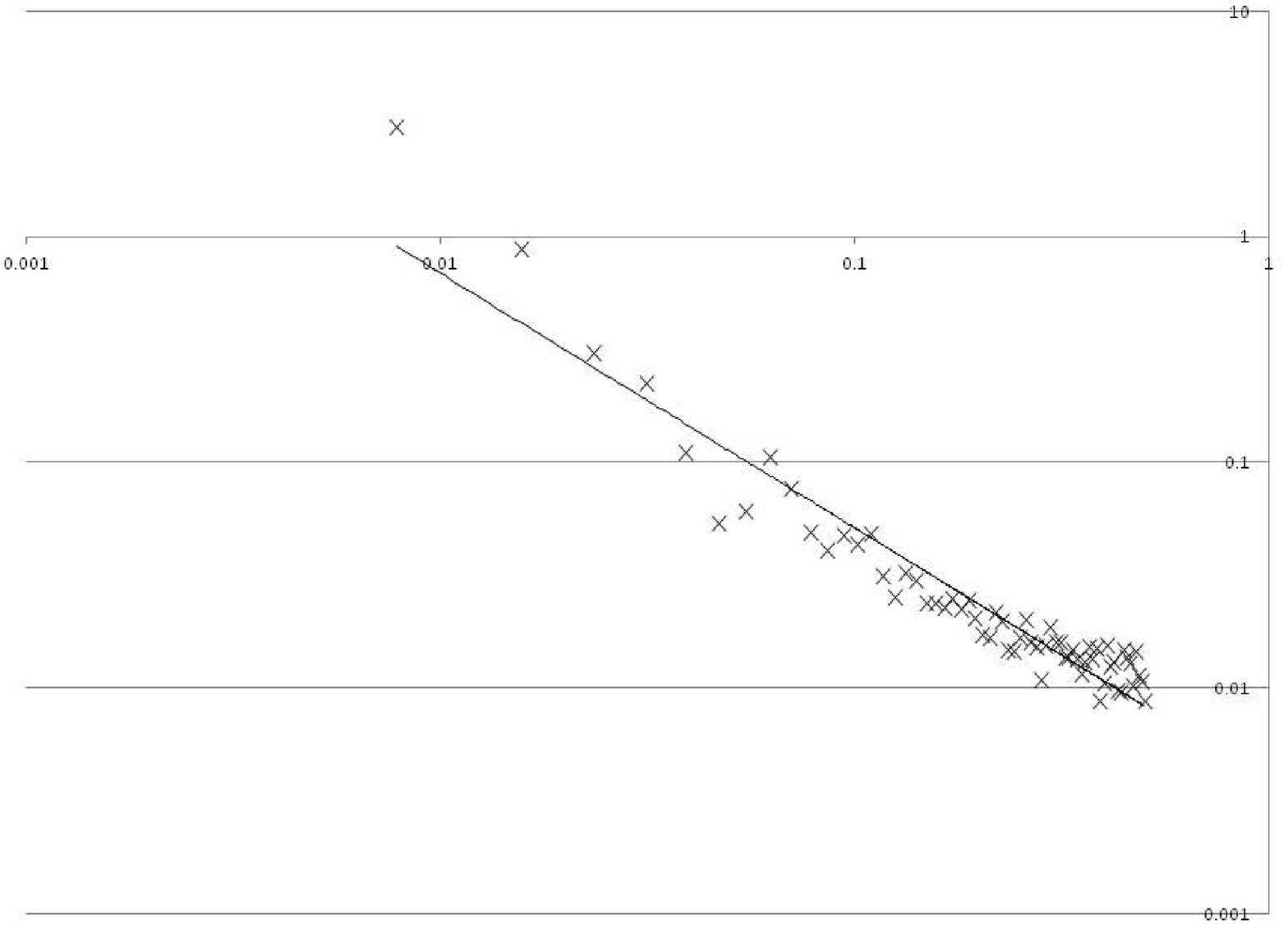} &
(b)\includegraphics[width=4.5cm]{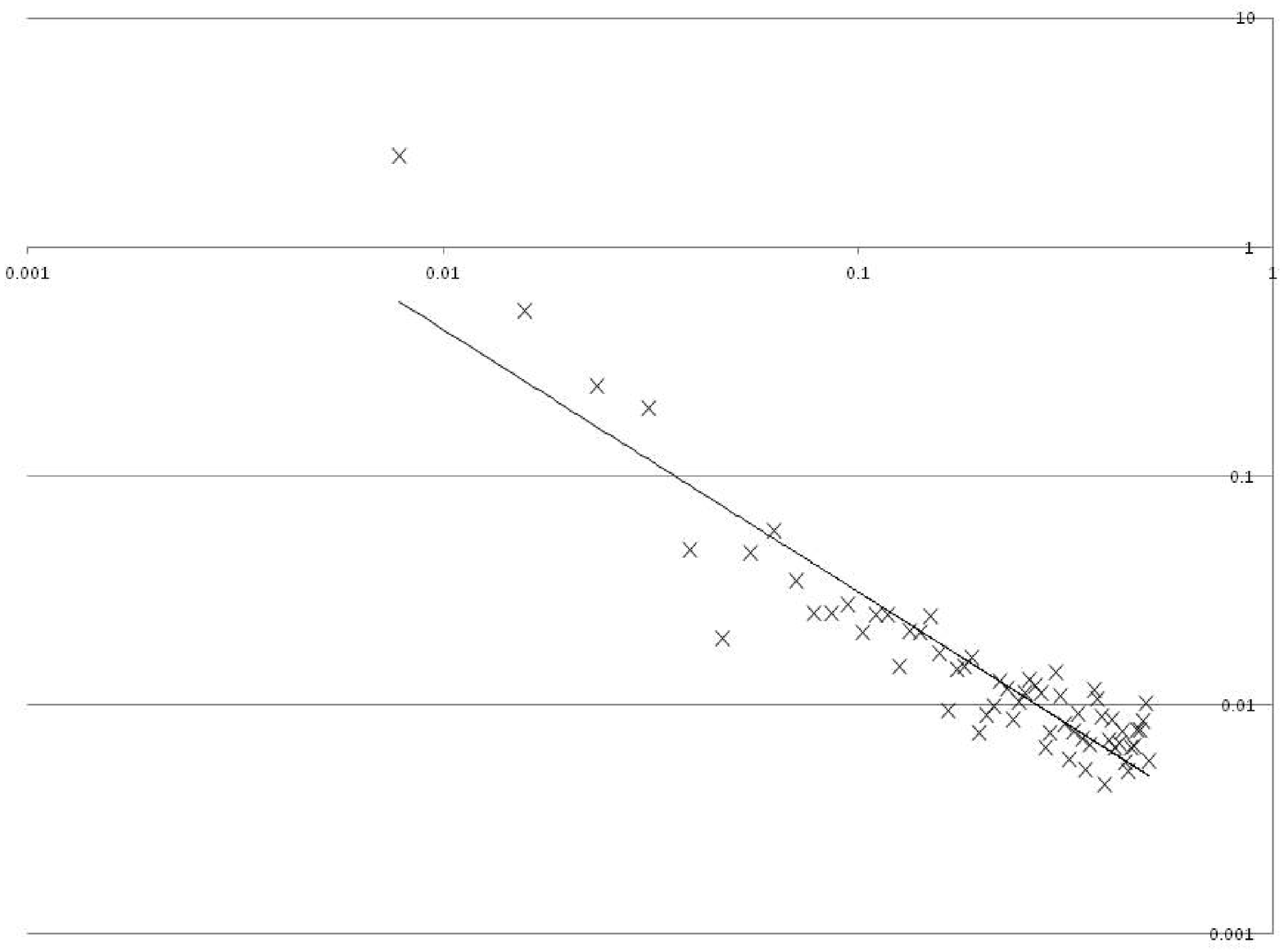} &
(c)\includegraphics[width=4.5cm]{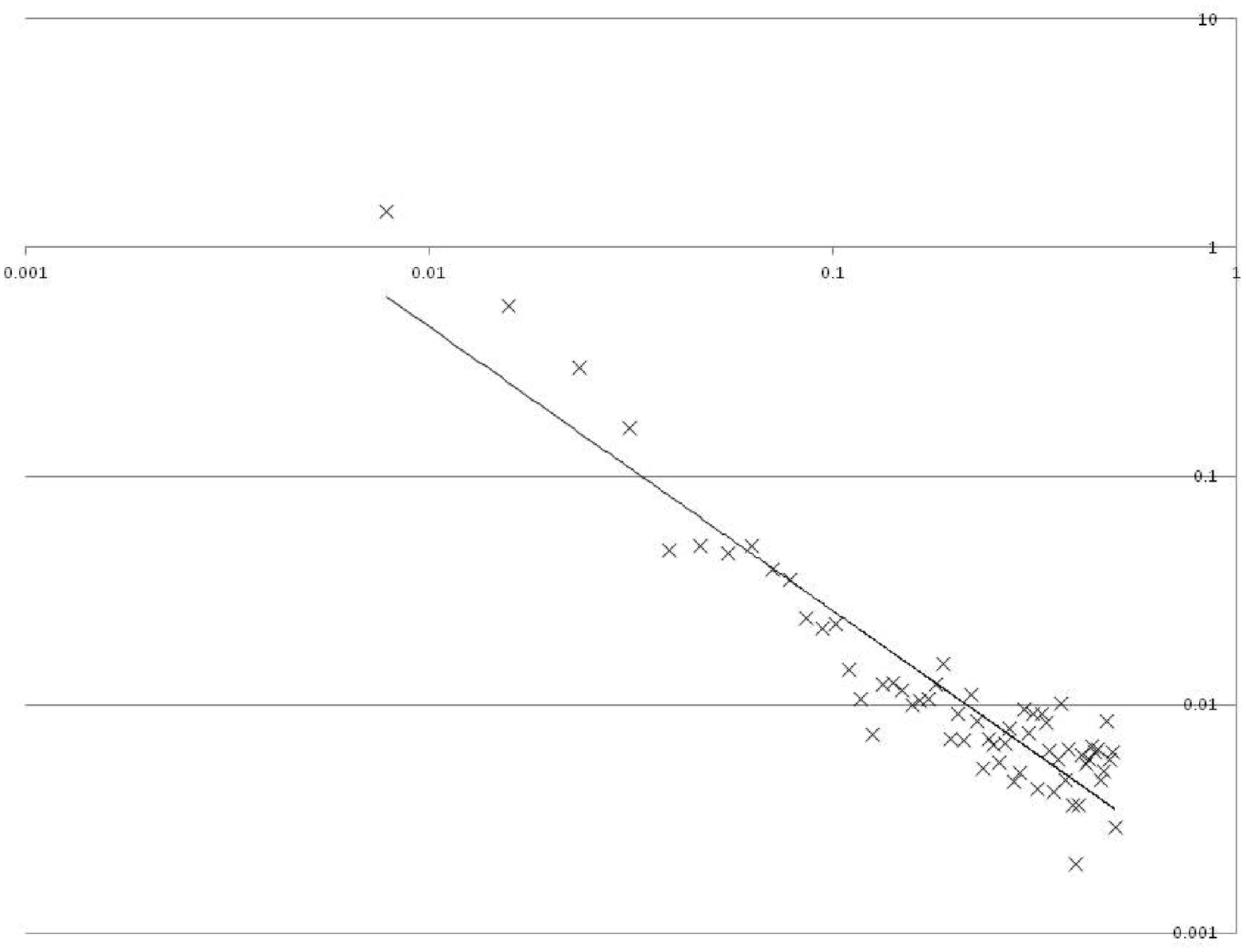} \\
(d)\includegraphics[width=4.5cm]{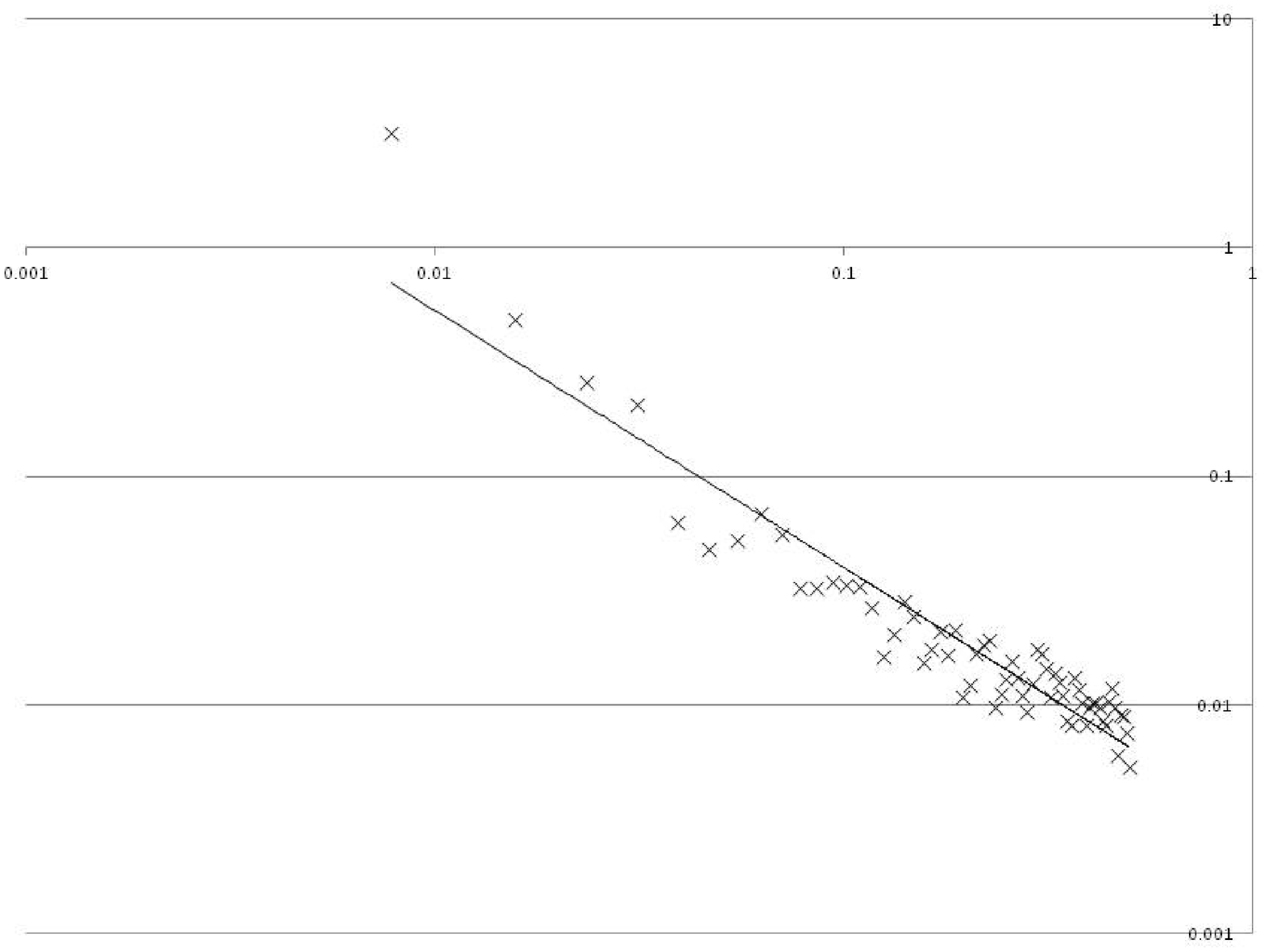} &
(e)\includegraphics[width=4.5cm]{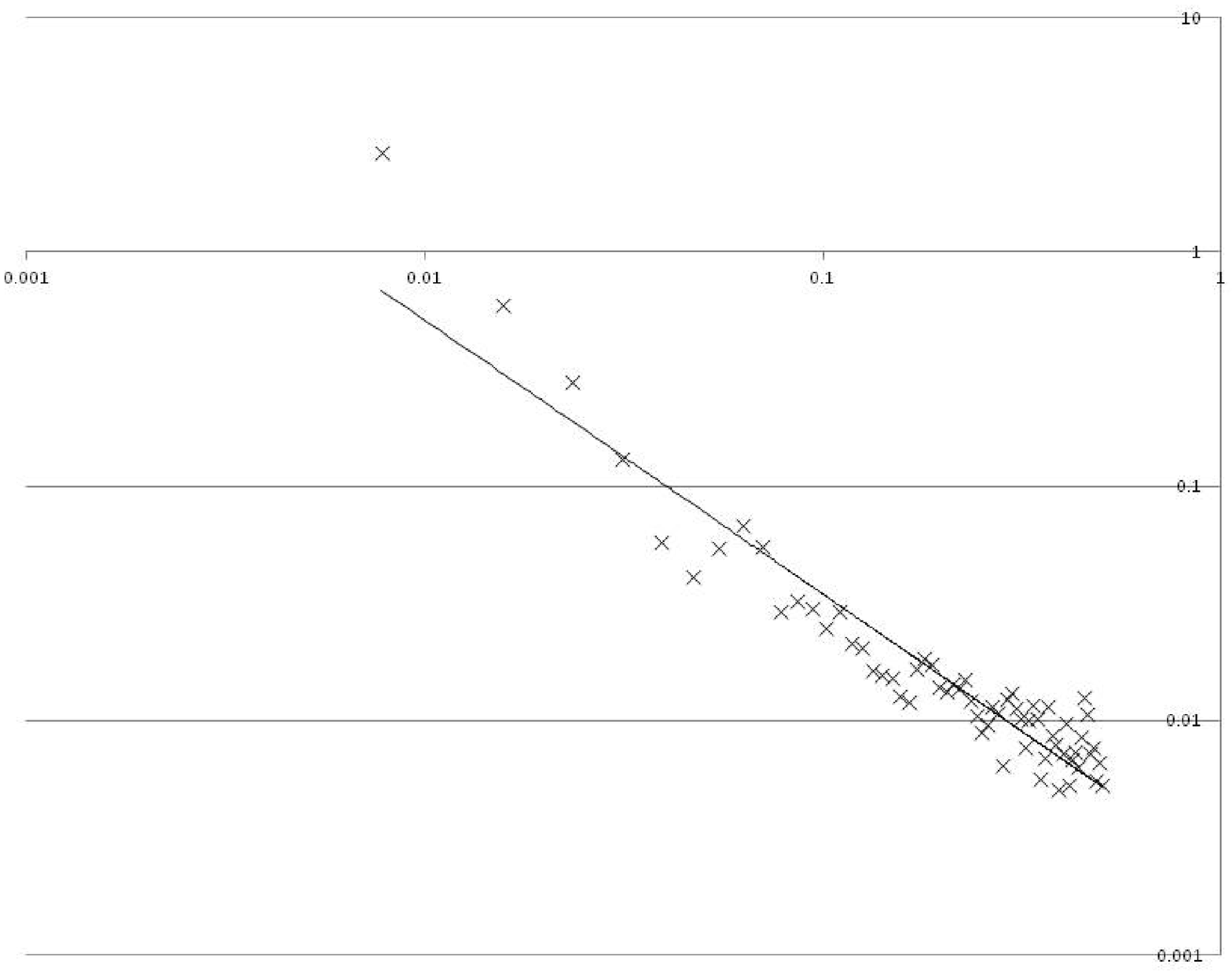} &
(f)\includegraphics[width=4.5cm]{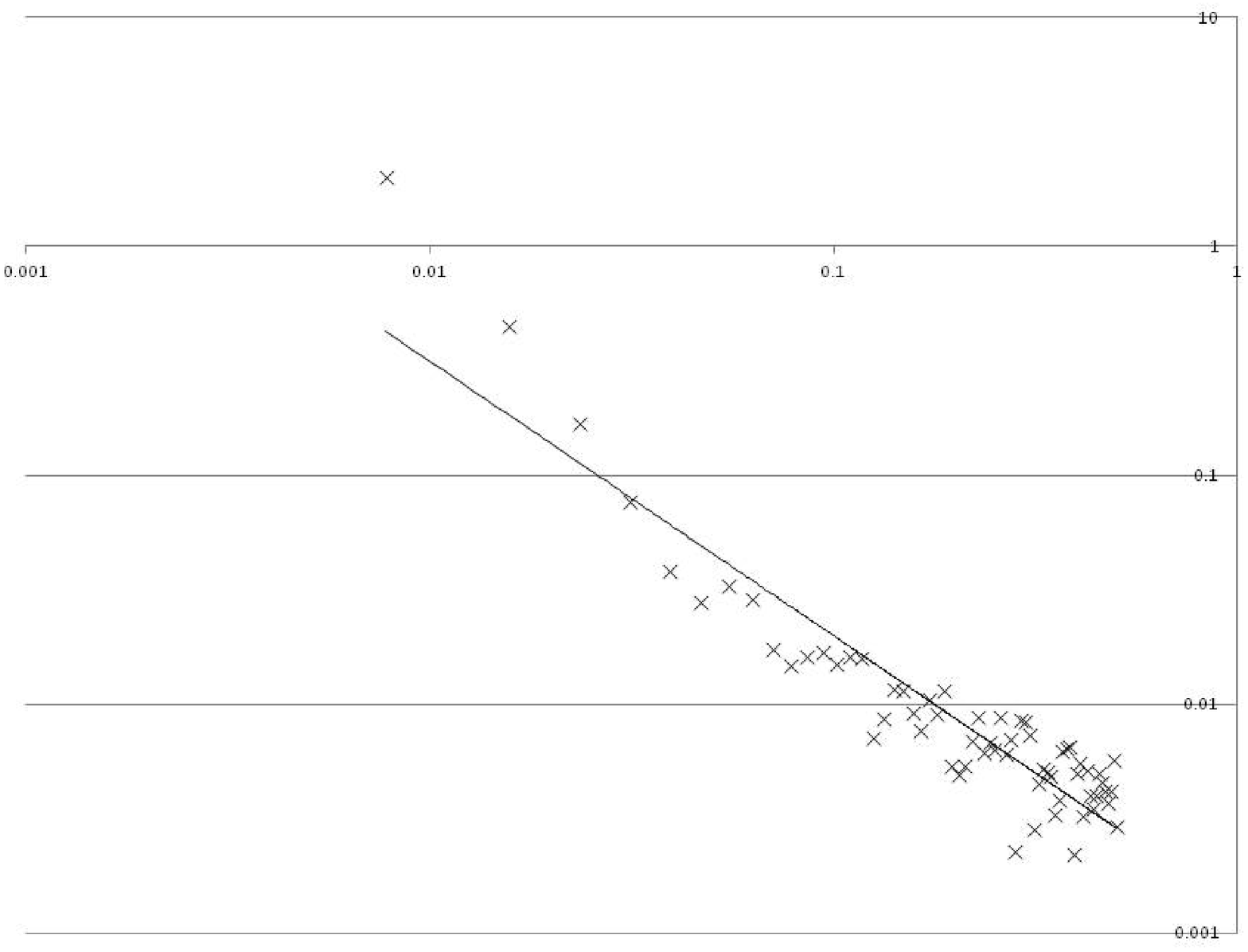} \\
\end{tabular}  
\caption{Plots of the spectrum analysis for (a) construction, (b) wholesale, (c) real estate, (d) retail sale, (e) other services, and (f) manufacturing.}
\label{7}
\end{figure}

Next, we apply our data to the risk data bank (RDB) data \cite{Data3}. The data covers all of the enterprise data without individual owner-managers in Japan.
The data is monthly from 2001 to 2017 and the seasonal effects were adjusted.
 The historical data and autocorrelation are shown in Fig. \ref{6} (a), which is different from the previous two samples.
The slow decay of the correlation was confirmed.
In Fig. \ref{6} (b) 1/f fluctuations were confirmed.
This corresponds to the power decay of the Wiener-Khinchin theorem, which shows the relationship between the autocorrelation and power spectrum by a Fourier transformation.
In Fig. \ref{7} we show the power spectrum for each sector, namely, construction, wholesale, real estate, retail sales, other services, and manufacturing.
The solid line represents the trend.
We can conclude that the temporal correlation may contain a long memory for this data.
However, it is difficult to confirm a strict power law.

\section{Estimation of parameters}
We estimate the long run probability of default, $\theta$, and the default correlation, $\rho_D$, for S\&P and Moody's data by the MAP estimation.
We use a uniform distribution for the prior distribution
$f(\theta,\rho_D)$.
As discussed in the previous section, the exponential and power decays are used for the temporal correlation.
The conclusions are listed in Table \ref{game3} for the exponential and
 power decay models.
We confirmed a small $r$ value that represents the small temporal correlation.
 The parameter $\gamma$ for the power decay is greater than the phase transition point, 
$\gamma=1$. 
The PD and default correlation are almost the same as the estimations by the exponential and power decay models. The reason is that the power index, $\gamma$, is adequately large and there is only a small difference between the exponential and power decay models.
The first- and second-year’s temporal correlations, $d_1$ and $d_2$, respectively, are important for representing the data.

The parameters depend on the data terms. In the recent past, the default and temporal correlations have become minimal.  
This may depend on the smooth financial operations of governments and central banks.
Alternatively, the long history data of ~100 years have long correlations that are less than the phase transitions.
This depends on the old data before the 1980s.
For the RDB data, we can estimate $\gamma=2$, which is in the normal convergence phase. 
Hence, we can estimate the PD by the Bayesian formula, which we 
introduced.
\begin{table}[tbh]
\caption{MAP estimation of the parameters for the exponential and power decay models. }
\begin{center}
\begin{tabular}{|l|l|ccc|ccc|r|}
\multicolumn{4}{c}{}\\ \hline
&  & &Exponential decay  &&&Power decay  & \\
No.& Model & 
 $\theta$& $\rho_D$ &$r$& $\theta$& $\rho_D$ &$\gamma$\\ \hline \hline
1&Moody's 1920-2017& 0.96\%& 1.9\% &0.044& 0.95\%& 2.0\% &4.7\\ \hline
2&Moody's 1920-2017 SG & 2.37\% & 3.9\% & 0.044& 2.35\% & 4.1\% & 4.7  \\ \hline
3& Moody's 1981- 2017& 1.49\% & 0.7\%&0.023& 1.46\% & 0.7\%&5.9 \\ \hline
4&Moody's 1990-2017& 1.65\% & 0.7\%& 0.006& 1.70\% & 0.8\%& 7.0  \\ \hline
5&Moody's 1981-2017 SG&4.25\% & 1.8\% & 0.020 &4.29\% & 1.8\% & 6.0 \\ \hline
6&S\&P 1981-2017& 1.54\% & 0.8\% & 0.024& 1.54\% & 0.8\% & 5.7  \\ \hline
7&S\&P 1990-2017& 1.72\% & 0.8\% & 0.006& 1.72\% & 0.8\% & 7.5  \\ \hline
8&S\&P 1990-2017  SG& 4.21\% & 2.0\% & 0.024& 4.17\% & 1.9\% & 5.7  \\ \hline
\end{tabular}
\label{game3}
\end{center}
\end{table}

\section{Concluding Remarks}

In this paper, we introduced a hierarchical Bayesian estimation method using the beta-binomial distribution to estimate the parameters, probability of default (PD), and default correlation.
Moreover, we considered a multi-year case with temporal correlation.
We confirmed phase transitions when the temporal correlation decayed by a power curve, which means that the correlation had a long memory.
Conversely, for the case of exponential decay, there was no phase transition.
When the power index, $\gamma$, was above or equal to one, the
estimator distribution of the PD converged.
Conversely, when the power index was less than 1, the distribution did not converge.
The critical exponent $0<\delta<1$ depended on the microscopic feature of the model and
the universality class of the phase transition differed from those
of the nonlinear P\'{o}lya urn.
We call this phase transition a "short memory-long memory transition".
In summary, the condition for the estimation of parameters is $\hat{T}=\lim_{\tau\rightarrow \infty} \int_{1}^{\tau}d(\tau-s)\textrm{d}s<\infty$.

To confirm the form of the decay, we investigated the empirical default history data using a Fourier transformation. We determined that the power spectrum of the default history was seemingly 1/f, which implies that the correlation had a long memory for the RDB monthly data.
We applied this method to the historical data and estimated the parameters.
The region of the power index provided normal convergence. We have demonstrated that, for adequate data collection, these parameters can be estimated correctly.

\begin{acknowledgments}
  This work is
  supported by JPSJ KAKENHI [Grant No. 17K00347]. 
\end{acknowledgments}

\appendix

\section{MAP estimation for Multi-year case}

We extend the maximum a posteriori (MAP) estimation, which we discussed in section 2 for the multi-year case.
The number of obligors and defaults in the $j^{th}$ year are $n_j$ and $k_j$, respectively.
 When a prior function, $f(\theta,\rho_D )$, is a constant function, the maximum point is
\begin{eqnarray}
\frac{\partial P(\theta,\rho_D|n_1,\cdots,nT, k_1, \cdots,k_T)}{\partial \theta}&\propto&
\frac{(1-\rho_D)}{\rho_D}\frac{\prod_{j=1}^{T}\Gamma(\alpha_j+k_j)}{\prod_{j=1}^{T}\Gamma(\alpha_j)}\frac{\prod_{j=1}^{T}\Gamma(n_j+\beta_j-k_j)}{\prod_{j=1}^{T}\Gamma(\beta_j)}
\nonumber \\&\times &
(\sum_{j=1}^{T}\{\varphi(\alpha_j+k_j)-\varphi(\alpha_j)-\varphi(\beta_j+n_j-k_j)+\varphi(\beta_j)\})
\nonumber \\
&=&
\frac{(1-\rho_D)}{\rho_D}\frac{\prod_{j=1}^{T}\Gamma(\alpha_j+k_j)}{\prod_{j=1}^{T}\Gamma(\alpha_j)}\frac{\prod_{j=1}^{T}\Gamma(n_j+\beta_j-k_j)}{\prod_{j=1}^{T}\Gamma(\beta_j)}
\nonumber \\
&\times&
\{\sum_{j=1}^{T}(\sum_{i=1}^{k_j}\frac{1}{\alpha_j+i-1}-\sum_{i=1}^{n_j-k_j}\frac{1}{\beta_j+i-1})\}=0,
\label{MAP}
\end{eqnarray}
where $\varphi(x)$ is the digamma function.
$\alpha_j$ and $\beta_j$ are the adjusted $\alpha$ and $\beta$. 
$\alpha_j=\alpha+\sum_{l=1}^{j-1}d_{j-l}k_l$ and 
$\beta_j=\beta+\sum_{l=1}^{j-1}d_{j-l}(n_l-k_l)$.
The first term in the last set of parentheses in Eq. (\ref{MAP}) is a monotonously decreasing function about $\theta$, because $\alpha$ increases.
The second term in the last set of parentheses is a monotonously increasing function about $\theta$, because $\beta$ decreases.
When $\theta\sim 0$, the difference of the two terms is positive because $\alpha_1=\alpha$.
In contrast, when $\theta\sim 1$, the difference of the two terms becomes negative because $\beta_1=\beta$.
Hence, the function, $P(\theta|X=k,\rho_D)$, has one peak in the range $0<\theta<1$.

\section{Scaling functions $f_{\xi}(\xi_t)$ and $f_{\tau}(\xi_t)$}
We define the relaxation and second-moment correlation times, $\tau(t)$ and 
$\xi(t)$, respectively, using the $n^{th}$ moment of $C(t)$ as in Eq. (\ref{moment}).
If we assume that $C(t)\propto t^{-\delta}$, $M_{n}(t)$
behaves as
\[
M_{n}(t)\propto
\left\{
\begin{array}{cc}
\frac{1}{n+1-\delta}t^{n+1-\delta} & \delta< n+1,  \\
\ln t  & \delta=n+1, \\
\frac{1}{\delta-(n+1)}  & \delta >n+1.
\end{array}
\right.
\]
Using the asymptotic behavior of $M_{n}(t)$, we find
$\tau(t)$ behaves as
\[
\tau(t)\propto
\left\{
\begin{array}{cc}
\frac{1}{1-\delta}t^{1-\delta} & \delta<1,  \\
\ln t  & \delta=1, \\
\mbox{constant}  & \delta>1.
\end{array}
\right.
\]
$\xi(t)$ behaves as
\[
\xi(t)\propto
\left\{
\begin{array}{cc}
\sqrt{\frac{1-\delta}{3-\delta}}t & \delta<1,  \\
t/\sqrt{\ln t} & \delta=1,  \\
\sqrt{\frac{\delta-1}{3-\delta}}t^{(3-\delta)/2}  & 1<\delta<3, \\
\mbox{constant}  & \delta\ge 3.
\end{array}
\right.
\]
The scaling function for $\tau$ is defined as
$f_{\tau}(\xi_t)\equiv \lim_{t\to\infty}\frac{\tau(st)}{\tau(t)},s>1$.
From the asymptotic behavior of
$\tau(t)$, we have
\[
f_{\tau}(\xi_t)\equiv \lim_{t\to\infty}\frac{\tau(st)}{\tau(t)}
=\left\{
\begin{array}{cc}
s^{1-\delta}  & 0<\delta<1  \\
1  & \delta\ge 1
\end{array}
\right.
\]
For $\delta<1$, $\xi_{t}\equiv \lim_{t\to\infty}\xi(t)/t
=\lim\sqrt{(1-\delta)/(3-\delta)}$ and
the scaling function is given in terms of $\xi_{t}$ as
\[
\log_{s}f_{\tau}(\xi_t)=1-\delta=\frac{2(\xi_t)^2}{1-(\xi_t)^2}.
\]
$\xi_t=1/\sqrt{3}$  and $f_{\tau}(\xi_t)=2$
in the limit $\delta\to 0$.

The scaling function for $\xi$ is defined as
$f_{\xi}(\xi_{t})\equiv \lim_{t\to\infty}\frac{\xi(st)}{\xi(t)}$. We have
\[
f_{\xi}(\xi_t)\equiv \lim_{t\to\infty}\frac{\xi(st)}{\xi(t)}=
\left\{
\begin{array}{cc}
s  & \delta \le 1  \\
s^{(3-\delta)/2}  & 1<\delta<3 \\
1  & \delta \ge 3
\end{array}
\right.
\]
By the renormalization transformation $t\to s^n t$,
$\lim_{n\to \infty}\xi(s^nt)/s^n=\xi(t)$ for $\delta\le 1$.
For $\delta>1$, $\xi(s^nt)/s^n=0$.
The critical state of the system exists at $\delta<1$.

We assume $C(t)\simeq c+\Delta C(t), c>0$ and
$\Delta C(t)$ rapidly decays to zero.
$\lim_{t\to \infty}\tau(t)=ct$
and $\xi_t=1/\sqrt{3}$.
$f_{\xi}(\xi_t)\equiv \lim_{t\to\infty}\xi(st)/\xi(t)=s$ and $
f_{\tau}(\xi_t)\equiv \lim_{t\to\infty}
\tau(st)/\tau(t)=s$ holds.




\begin{thebibliography}{1} 
\bibitem{Bro} D. Brockmann, L. Hufinage, and Geisel, Nature {\bf 439}, 462 (2006).
\bibitem{W2} I. T. Wong, M. L. Gardel, D. R. Reichman, E. R. Weeks, M. T. Valentine, A. R. Bausch, and D. A. Weitz, Phys.
Rev. Lett. {\bf 92}, 178101 (2004). 
 \bibitem{G} Y. Gefen Y., A. Aharony, and S. Alexander S., Phys. Rev.
Lett. {\bf 50}, 77 (1983). 
\bibitem{M} R. Metzler and J. Klafter, Phys. Rep. {\bf 339}, 1 (2000).
\bibitem{galam} G. Galam, Stat. Phys. {\bf 61}, 943 (1990).
\bibitem{galam2} G. Galam, Inter J. Mod. Phys. C
{\bf 19(03)}, 409 (2008).
\bibitem{Hisakado7} M. Hisakado, F. Sano, and S. Mori, Phys Soc. Jpn {\bf 87 (2)}, 024002 (2018).
\bibitem{Man} N. M. Mantegna and H. E. Stanley, {\it Introduction to Econophysics: Correlations and Complexity in Finance}
(Cambridge University Press, 2000).
\bibitem{Mori} Mori S. and Hisakado M., J. Phys. Soc. Jpn. {\bf 79}, 034001 (2010).
\bibitem{Hisakado2} M. Hisakado and S. Mori, J. Phys. A {\bf 43}, 31527 (2010).
\bibitem{Hisakado3} M. Hisakado and S. Mori, J. Phys. A {\bf 44}, 275204 (2011).
\bibitem{Hisakado4} M. Hisakado and S. Mori, Physica A {\bf 417}, 63 (2015).
\bibitem{Hisakado5} M. Hisakado and S. Mori, Phys. A. {\bf 108}, 570 (2016).
\bibitem{Hod} S. Hod and U. Keshet, Phys. Rev. E {\bf 70}, 11006 (2004).
\bibitem{M2010} S. Mori, K. Kitsukawa, and M. Hisakado, Quant. Fin.{\bf 10}, 1469 (2010).
\bibitem{M2008} S. Mori, K. Kitsukawa, and M. Hisakado, J. Phys. Sco.Jpn. {\bf 77}, 114802 (2008).
\bibitem{Sch} P. J. Sch\"onbucher, {\it Cresit Derivatives Pricing Models:Models, Pricing, and Inplementation} (John Wiley \& Sons, Ltd. 2003).

\bibitem{Tas} K. Pluto and D. Tasche,
{\it Estimating Probabilities of Default for Low Default Portfolios}
In: Engelmann B., Rauhmeier R. (eds) The Basel II Risk Parameters. Springer, Berlin, Heidelberg (2011). 
\bibitem{FSA} N. Benjamin, A. Cathcart, and K. Ryan K,{\it Low Default Portfolios: A Proposal for Conservative Estimation of Default Probabilities} (Financial Services Authority, 2006).
\bibitem{Hisakado} M. Hisakado, K. Kitsukawa, S. and Mori, J. Phys. A {\bf 39}, 15365 (2006).
\bibitem{Wit} G. Witt, {\it Moody's Coorelated Binomial default distribution} (Moody's, 2006). 
\bibitem{Mer} R. C. Merton, J. Fin. {\bf 29(2)}, 449 (1974).
\bibitem{Mori5} S. Mori and M. Hisakado, Phys Rev. E
{\bf 92}, 052112 (2015).
\bibitem{Mori6} S. Mori and M. Hisakado, J. Phys. Soc. Jpn. {\bf 84}, 054001 (2015).  
\bibitem{Kes} M. S. Keshner, Proc. IEEE. {\bf 70}, 212 (1982).
\bibitem{AR} R. F. Engle, Econometarica {\bf 50(4)}, 1912773 (1982).
\bibitem{GA} T. Bollerslev, J. Econometrics {\bf 31(3)}, 307 (1986).
\bibitem{Mori18} S. Mori, M. Hisakado, and K. Nakayama, Mean field vector model on networks and multi-variate beta distribution
arXiv preprint arXiv:1810.05643 (2018).
\bibitem{Polya} G. Polya, Ann. Inst. Henri Poincar\'{e} {\bf 1}, 117 (1931).
\bibitem{Data1} 
{\it 2016 Annual Global Corporate Default Study and Rating Transitions} (Standard \& Poor's Rating Services, 2017).

\bibitem{Data2} 
{\it Moody's Annual Default Study: Corporate default and recovery dates, 1920-2017} (Moody's 2018).

\bibitem{Hec} W. B. Hickman,{\it Corporate and Investor Experience}
(Princeton University Press, 1958).


\bibitem{Data3} Risk Data Bank 2018,{\it https://www.riskdatabank.co.jp/rdb/top/}





\end{thebibliography}
\end{document}